\documentclass{aa}  
\usepackage{graphicx}
\usepackage{txfonts}
\usepackage{siunitx}  
\DeclareSIUnit\parsec{pc}
\usepackage{csquotes}
\usepackage{subcaption}
\usepackage{caption}
\usepackage{hyperref}
\hypersetup{colorlinks=true,linkcolor=blue,citecolor=blue,filecolor=blue,urlcolor=blue}
\usepackage{orcidlink}

\begin{document}

   \title{From gas to stars along the spiral wave: CO, HCN, and star\\ formation variations across the spiral arms in NGC\,4321 and M51}

   \author{Minou~Greve\orcidlink{0009-0006-6310-2583}
          \inst{\ref{AIfA}}\fnmsep\thanks{e-mail: mgreve@astro.uni-bonn.de}
          \and Lukas~Neumann\orcidlink{0000-0001-9793-6400}\inst{\ref{ESO}}
          \and Mallory~Thorp\orcidlink{0000-0003-0080-8547}\inst{\ref{AIfA}}
          \and Dario~Colombo\orcidlink{0000-0001-6498-2945}\inst{\ref{AIfA}}
          \and Frank~Bigiel\orcidlink{0000-0003-0166-9745}\inst{\ref{AIfA}}
          \and Miguel~Querejeta\orcidlink{0000-0002-0472-1011}\inst{\ref{OAN}}
          \and Sharon~E.~Meidt\orcidlink{0000-0002-6118-4048}\inst{\ref{UGent}}
          \and Ashley~T.~Barnes\orcidlink{0000-0003-0410-4504}\inst{\ref{ESO}}
          \and Zein~Bazzi\orcidlink{0009-0001-1221-0975}\inst{\ref{AIfA}}
          \and Ralf~S.~Klessen\orcidlink{0000-0002-0560-3172}\inst{\ref{ITA},\ref{IWR}}
          \and Adam~K.~Leroy\orcidlink{0000-0002-2545-1700}\inst{\ref{Ohio}}
          \and Hsi-An~Pan\orcidlink{0000-0002-1370-6964}\inst{\ref{TKU}}
          \and Jérôme~Pety\orcidlink{0000-0003-3061-6546}\inst{\ref{IRAM},\ref{LUX}}
          \and
          Marina~Ruiz-Garc\'ia\orcidlink{0000-0002-8305-1801}\inst{\ref{OAN}, \ref{UCM}}
          \and Eva~Schinnerer\orcidlink{0000-0002-3933-7677}\inst{\ref{MPIA}}
          \and Rowan~Smith\orcidlink{0000-0002-0820-1814}\inst{\ref{StA}}
          \and Sophia~Stuber\orcidlink{0000-0002-9333-387X}\inst{\ref{Japan}}\fnmsep\inst{\ref{MPIfR}}\thanks{International Research Fellow of Japan Society for the Promotion of Science (Postdoctoral Fellowships for Research in Japan)}
          \and Jiayi~Sun\orcidlink{0000-0003-0378-4667}\inst{\ref{Kentucky}}
          \and Antonio~Usero\orcidlink{0000-0003-1242-505X}\inst{\ref{OAN}}
          \and Thomas~G.~Williams\orcidlink{0000-0002-0012-2142}\inst{\ref{JBCA}}
          }

   \institute{Argelander-Institut für Astronomie, Universität Bonn, Auf dem Hügel 71, 53121 Bonn, Germany\label{AIfA}
   \and
   European Southern Observatory, Karl-Schwarzschild Straße 2, 85748 Garching bei München, Germany\label{ESO}
   \and 
   Observatorio Astron\'omico Nacional (IGN), C/ Alfonso XII, 3, E-28014 Madrid, Spain\label{OAN}
   \and
   Sterrenkundig Observatorium, Universiteit Gent, Krijgslaan 281 S9, B-9000 Gent, Belgium\label{UGent}
   \and
   Universit\"{a}t Heidelberg, Zentrum f\"{u}r Astronomie, Institut f\"{u}r Theoretische Astrophysik, Albert-Ueberle-Str.\ 2, 69120 Heidelberg, Germany\label{ITA} 
   \and
   Universit\"{a}t Heidelberg, Interdisziplin\"{a}res Zentrum f\"{u}r Wissenschaftliches Rechnen, Im Neuenheimer Feld 225, 69120 Heidelberg, Germany\label{IWR}
   \and
   Department of Astronomy, The Ohio State University, 140 West 18th Avenue, Columbus, OH 43210, USA\label{Ohio}
   \and
   Department of Physics, Tamkang University, No.151, Yingzhuan Road, Tamsui District, New Taipei City 251301, Taiwan\label{TKU}
   \and 
   IRAM, 300 rue de la Piscine, 38400 Saint Martin d'H\`eres, France\label{IRAM}
   \and
   LUX, Observatoire de Paris, PSL Research University, CNRS, Sorbonne Universités, 75014 Paris, France\label{LUX}
   \and
   Facultad de Ciencias F\'{\i}sicas, Pl. de Ciencias 1, Universidad Complutense de Madrid, E-28040 Madrid, Spain\label{UCM}
   \and
   Max Planck Institute for Astronomy, Königstuhl 17, 69117 Heidelberg, Germany\label{MPIA}
   \and
   University of St Andrews, SUPA, School of Physics \& Astronomy, North Haugh, St Andrews, UK, KY16 9SS \label{StA}
   \and
   National Astronomical Observatory of Japan, 2-21-1 Osawa, Mitaka, Tokyo 181-8588, Japan\label{Japan}
   \and
   Max Planck Institute for Radio Astronomy, Auf dem Hügel 69, 53121 Bonn, Germany\label{MPIfR}   
   \and
   Department of Physics and Astronomy, University of Kentucky, 506 Library Drive, Lexington, KY 40506, USA\label{Kentucky}
   \and
   UK ALMA Regional Centre Node, Jodrell Bank Centre for Astrophysics, Department of Physics and Astronomy, The University of Manchester, Oxford Road, Manchester M13 9PL, UK\label{JBCA}
   }

   \date{Received 23 December 2025; accepted 18 April 2026}

  \abstract
    {Molecular clouds form stars from the interstellar medium via gravitational collapse, following a sequence from low-density gas to high-density cores and eventually the formation of stars. In classical density wave theory, gas clouds orbiting the galaxy experience gas compression and triggered star formation, while encountering the gravitational well of spiral arms.} 
   {We aim to trace these different phases of the molecular cloud life cycle via tracers of molecular gas (CO), dense molecular gas (HCN), and star formation (H$\alpha$, 24\,$\mu$m) within the spiral arms of two grand-design spiral galaxies: NGC\,4321 and M51 (NGC\,5194).} 
   {In the spiral arms of these galaxies, we investigate the relation between molecular gas, dense gas, and star formation (CO-HCN-SFR) at matched physical resolutions of 270\,pc and 125\,pc in NGC\,4321 and M51, respectively. We employed spiral arm masks for these galaxies and investigate trends of HCN/CO and SFR/HCN (SFR/CO), which serve as proxies for the dense gas fraction and dense (molecular) gas star formation efficiency, perpendicular to the spiral arm spines.} 
   {We find that HCN/CO, SFR/CO, and SFR/HCN increase from the upstream towards the downstream side of both spiral arms of NGC\,4321, while their trends are less prominent in M51.}
   {Our results indicate that large-scale galactic dynamics (e.g. density waves) can induce a sequence of gas density and star formation-to-gas density variations perpendicular to the spiral arms. This sequence contributes to the increased scatter seen among spectroscopic ratios such as HCN/CO and SFR/HCN at sub-kiloparsec scales.}

    \titlerunning{CO, HCN, and star formation variations across the spiral arms in NGC\,4321 and M51}
   \maketitle

\section{Introduction}
Molecular gas is the fuel for star formation in galaxies, where stars form primarily in the dense clumps of giant molecular clouds (GMCs; \citealt{McKee_2007, Bigiel_2008, Klessen_2016}). However, the question of how star formation proceeds in the dense molecular gas in external galaxies is not well understood. This is partly because higher-density gas tracers such as HCN(1--0), HCO$^+$(1--0), CS(2--1), and N$_2$H$^+$(1--0) are significantly fainter than the commonly observed low-J CO transitions, which trace the bulk molecular gas at densities of $n(\mathrm{H}_2)\sim 10^2-10^3$\,cm$^{-3}$. Nevertheless, the tight correlation between HCN luminosity, a proxy for dense molecular gas \citep[with an effective excitation density $n_\textrm{eff}\sim10^4$\,cm$^{-3}$;][]{Shirley_2015}, and star formation rate (SFR) suggests the existence of a universal dense gas-to-star formation relation; in other words, a fixed dense gas star formation efficiency (SFE$_\mathrm{dense}$), defined as the SFR per unit dense gas mass \citep[the Gao-Solomon relation;][]{Gao_Salomon_2004}. This comprehensive relation between dense gas and star formation has been studied across many systems \citep[e.g.][]{Neumann_2025}.
However, over the past decade, resolved studies of nearby galaxies at kiloparsec (kpc) scales have found systematic variations of SFR/HCN, a proxy of SFE$_\mathrm{dense}$, within galaxies as a function of physical environment \citep{Usero_2015, Bigiel_2016, Jimenez_Donaire_2019, Neumann_2025} and molecular cloud conditions \citep{Gallagher_2018b, Neumann_2023a}. These studies have suggested that the physical conditions within galaxies affect how easily dense gas can form, as traced via the HCN/CO line ratio, a proxy of the dense gas fraction ($f_\mathrm{dense}$), as well as how efficiently this dense gas can be converted into stars \citep[see also review by][]{Schinnerer_Leroy_2024}. 

Recently, some of these nearby galaxies have been observed in dense gas tracers at even higher spatial resolution with ALMA \citep[NGC\,4321 at 270\,pc;][]{Neumann_2024} and NOEMA \citep[M51 at 125\,pc;][]{Stuber_2023}. These studies suggest that dynamical effects such as streaming motions, shear, or cloud-cloud collisions (e.g. across galactic bars) can additionally drive variations of HCN/CO and SFR/HCN that cannot be explained by the structural variations (i.e. ISM pressure, cloud surface density, or gas velocity dispersion) mentioned above. Instead, these variations might reflect the influence of galactic dynamics on the build up of dense gas in clouds and its ability to form stars. In particular, large-scale stellar bars and spiral arms may be influencing gas structure and also coordinating where and when star formation occurs.  

In the traditional density wave picture, spiral arms are rigidly rotating features and, as a result of the high relative speed between the spiral and orbiting material, gas can develop an upstream shock as it interacts with the spiral \citep{Roberts_1969}. The density enhancement built by the shock leads to elevated star formation. Over time, the sites of recent star formation move further away from the present location of the shock \citep{Lubow_1986, Roberts_1987, Egusa_2009, Querejeta_2025}. Measurements of gas density contrasts traced by CO in PHANGS\footnote{Physics at High Angular resolution in Nearby GalaxieS; https://www.phangs.org} targets suggest that shocks and gas self-gravity are indeed influencing the development of density enhancements across spirals \citep{Meidt_2020, Querejeta_2024}. In several cases, the spatial offset between gas density (traced by CO line emission) and recent star formation (traced by H$\alpha$ emission or by young stellar clusters) predicted in the density wave picture has been observed \citep{Tamburro_2008, Foyle_2010, Kim_2023, Querejeta_2025}.  
To date, however, the evidence for an evolutionary sequence from gas to stars across spiral arms, from their upstream to their downstream sides, has not considered dense gas tracers, such as HCN(1--0) or HCO$^+$(1--0). These should be more closely linked to the immediate sites of star formation and hence uncover an intermediate step between bulk molecular gas (CO) and star formation (SFR).
In this work, we utilise recent high-resolution, sensitive dense gas tracer maps of two nearby, relatively face-on, grand-design spiral galaxies: NGC\,4321 \citep[$i= \SI{38.5}{\degree}$;][]{Lang_2020}, characterised by a strong bar, and M51 \citep[$i= \SI{22}{\degree}$;][]{Colombo_2014b}, which is interacting with its companion NGC\,5195 (see Table~\ref{tab:properties} for their main properties). These two galaxies provide the only $< 300\,$pc scale, sensitive dense gas tracer maps of grand-design spirals to date that are paired with ancillary molecular gas and star formation rate tracers.
For the first time, these observations, with an angular resolution several times finer than the spiral arm widths, have enabled us to investigate the molecular gas-dense gas-star formation (CO-HCN-SFR) sequence at 270\,pc and 125\,pc scales. In particular, we study these tracers and their ratios (HCN/CO, SFR/CO, and SFR/HCN) perpendicular to the spiral arms of these two spiral galaxies.

This paper is structured as follows. In Sect.~\ref{sect:data} we present a summary of the data of NGC\,4321 and M51 utilized in this work. Then, in Sect.~\ref{sect:method}, we introduce the method to measure offsets from the spiral arm spines based on the morphological environmental masks. In Sect.~\ref{sect:results}, we present our results, displaying spectroscopic line ratios (HCN/CO, SFR/CO, SFR/HCN) variations across the spiral arms of NGC\,4321 and M51, which are discussed and compared to previous studies in Sect.~\ref{sect:discussion}. Finally, we summarize the key findings in Sect.~\ref{sect:conclusion}.

\section{Data}\label{sect:data}
\begin{table}[]
    \begin{center}  
    \caption{Properties of the targets NGC\,4321 and M51.}
    \begin{tabular}{lcc}
    \hline\hline 
    Property & NGC\,4321 & M51 \\ \hline 
    Inclination, $i [\si{\degree}]$ & $(38.5 \pm 2.4)^\mathrm{{(a)}}$ & $(22\pm5)^\mathrm{(e)}$\\
    Distance, $ d$\,[Mpc] & $(15.21 \pm 0.49)^\mathrm{{(b)}}$ & ($8.58\pm0.10)^\mathrm{(f)}$\\
    Morphology & SAB(s)bc$^\mathrm{(c)}$ & SAbc$^\mathrm{(c)}$ \\
    SFR\,[M$_\odot$\,yr$^{-1}$] & $(3.56 \pm 0.92)^{\mathrm{(d)}}$ & $(4.47\pm2.05)^{\mathrm{(g)}}$\\
    $\log_{10}(M_* / $M$_\odot)$ & $(10.75\pm0.11)^\mathrm{{(d)}}$ & $(10.73 \pm 0.10)^{\mathrm{(g)}}$\\
    \hline\hline
    \end{tabular}\label{tab:properties}
    \end{center}

\tablefoot{
(a) \citet{Lang_2020}; (b) \citet{Anand_2021};
(c) NASA Extragalactic Database (NED);
(d) \citet{Leroy_2019}; (e) \citet{Colombo_2014b}; (f) \citet{McQuinn_2016}; (g) \citet{Leroy_2020}.}
\end{table}

\subsection{NGC\,4321} 
    In this work, we used the HCN(1--0) integrated intensity map (moment-0 map) presented in~\citet{Neumann_2024} to trace the dense molecular gas at a high angular resolution of $\SI{3.7}{\arcsecond}$ ($\sim\SI{270}{\parsec}$) in the nearby spiral NGC\,4321 \citep[$d=15.2$\,Mpc;][]{Anand_2021}. 
    We used the moment-0 map of CO(2--1), tracing the bulk molecular gas, which is based on PHANGS–ALMA observations \citep{Leroy_2021b} and was convolved from an original $\SI{1.67}{\arcsecond}$ ($\sim\SI{120}{\parsec}$) resolution to the HCN resolution as described in~\citet{Neumann_2024}. We note that we decided against using the archival ALMA CO(1--0) observations as a tracer of molecular gas due to their poorer angular resolution of $\SI{4}{\arcsecond}\sim\SI{300}{\parsec}$.
    We combined the moment-0 maps with a map of the star formation rate surface density ($\Sigma_\text{SFR}$) based on the attenuation-corrected H$\alpha$ flux density from PHANGS–MUSE \citep{Emsellem_2022}. For the analysis, we convolved the $\Sigma_\text{SFR}$ map from $\sim\SI{1.16}{\arcsecond}$ to the HCN(1--0) resolution (using \texttt{convolution.convolve} from \texttt{astropy}) and re-projected it onto the same pixel grid. From these integrated intensities, $W$, we determined the line ratio $W_\text{HCN(1--0)}$/$W_\text{CO(2--1)}$ (hereafter, HCN/CO) as a proxy of the dense gas fraction ($f_\text{dense}$), $\Sigma_\text{SFR}$/$W_\text{CO(2--1)}$ (hereafter, SFR/CO) as a proxy for the molecular gas star formation efficiency (SFE$_\text{mol}$) as well as $\Sigma_\text{SFR}$/$W_\text{HCN(1--0)}$ (hereafter, SFR/HCN) as a proxy for the dense gas star formation efficiency (SFE$_\text{dense}$).
\begin{figure*}
    \centering
    \includegraphics[width=1\linewidth]{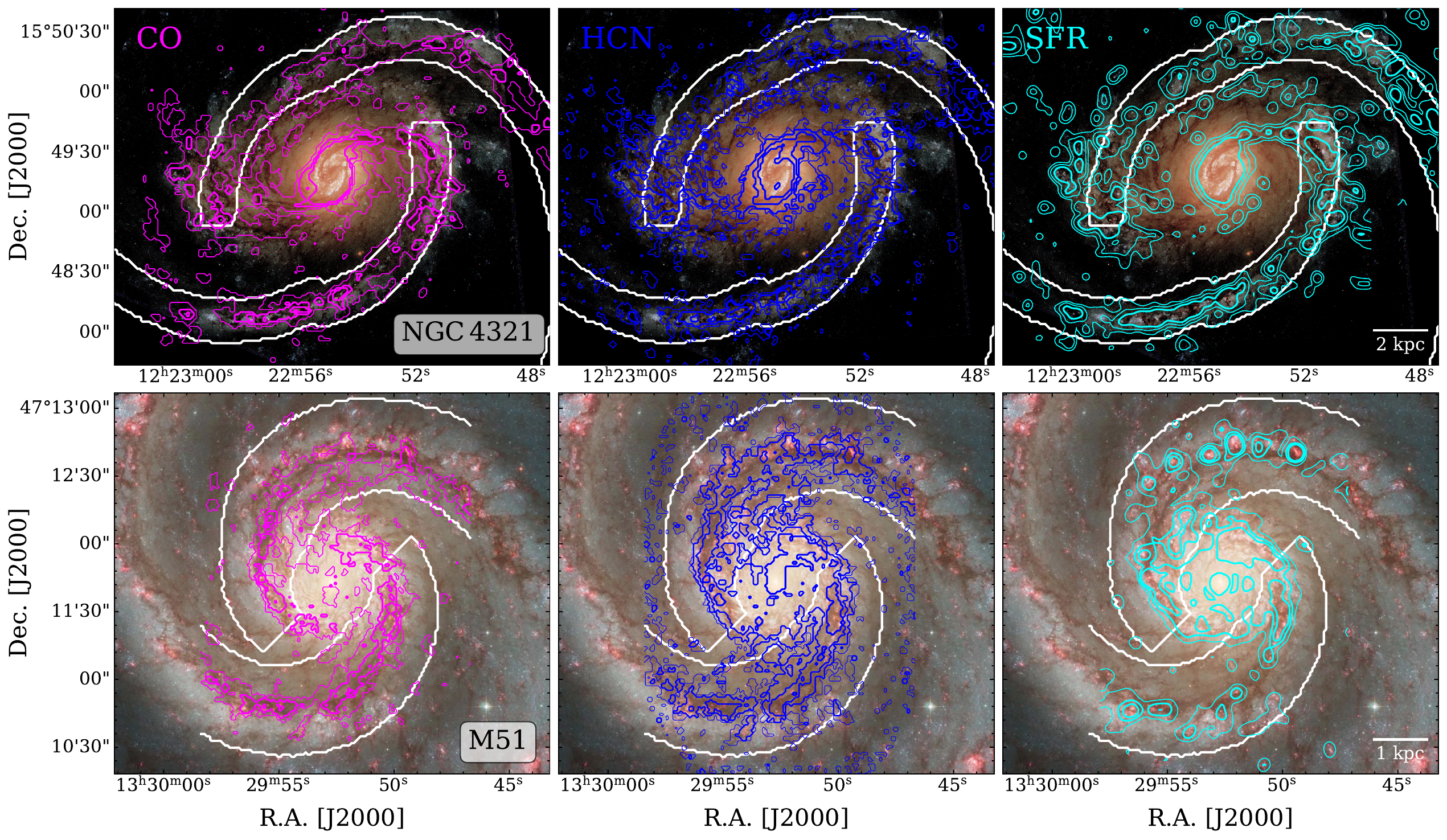}
    \caption{Optical \textit{Hubble} Space Telescope (HST) images of the targets NGC\,4321 (top) and M51 (bottom) overlaid with CO, HCN and SFR contours (from left to right). The CO (HCN) contours are drawn at S/N$ \geq 10, 30, $ and 60 (1, 3, 10, 30). The SFR contours correspond to the 70th, 85th, and 95th percentiles of the data. The spiral arm masks are indicated as white contours. Background HST image credit: NGC\,4321: PHANGS-HST \citep{PHANGS-HST}; M51: NASA, ESA, S. Beckwith (STScI) and the Hubble Heritage Team (STScI/AURA).}
    \label{fig:3x2_contours}
\end{figure*}

\subsection{M51}    
    We also studied the nearby \citep[$d=8.58$\,Mpc;][]{McQuinn_2016} grand-design spiral M51 by using the CO(1--0) data from the \enquote{PdBI Arcsecond Whirlpool Survey} \citep[PAWS;][]{Schinnerer_2013} in combination with the HCN(1--0) data from \enquote{Surveying the Whirlpool at Arcseconds with NOEMA}\,\citep[SWAN;][]{Stuber_2025a}. This IRAM Northern Extended Millimetre Array (NOEMA)+30\,m large program mapped the central $5\times\SI{7}{\kilo\parsec^2}$ of M51. We use the HCN(1--0) and CO(1--0) moment-0 FITS maps at a cloud-scale resolution of $\SI{3}{\arcsecond}\sim\SI{125}{\parsec}$, which were created with the PyStructure package \citep{pystructure}\footnote{https://github.com/PhangsTeam/PyStructure} to ensure comparability with \citet{Stuber_2025b}. We further include the $\Sigma_\text{SFR}$-map presented in \citet{Stuber_2025b}, which combines a Spitzer $\SI{24}{\micro\meter}$ map processed by \citet{Dumas_2011} with $\SI{3}{\arcsecond}$ resolution H$\alpha$ maps from \citet{Kessler_2020}.

\section{Method}\label{sect:method}
\subsection{Spiral masks}\label{sect:method_spiralmasks}

    To investigate the behaviour of $W_\text{CO}$, $W_\text{HCN}$, $\Sigma_\text{SFR}$, and their ratios across the spiral arms, we rely on the \enquote{simple} morphological environmental masks from \citet{Querejeta_2021}, shown in Fig.~\ref{fig:3x2_contours}. These are based on 3.6\,$\mu$m NIR photometry that traces stellar structures with a minimal impact of dust extinction. The spiral masks were created by (i) identifying bright 3.6\,$\mu$m features along each arm using an unsharp-mask technique and fitting log-spiral functions in the galaxy plane, (ii) assigning empirically determined widths based on CO emission, and (iii) visually refining the masks to ensure continuity. The NIR data and fits of both NGC\,4321 and M51 come from S$^4$G \citep{Herrera-Endoqui_2015}. These masks comprise smooth log-spiral segments with typical widths of the spiral arms of $\sim1{-}2$\,kpc, sufficient to cover most 3.6\,$\mu$m, CO, and H$\alpha$ emission associated with spiral arms despite local irregularities. 
    In the provided mask of NGC\,4321, the southern spiral arm splits into two branches beyond $\Delta\phi_6\sim15$\,kpc. To simplify the analysis, we neglect the inner branch after the split. This does not affect our results because there are no HCN detections in this segment. As M51 was not part of the public release from \citet{Querejeta_2021}, we constructed an analogous mask (Querejeta et al.\ in prep). 

\begin{figure}
    \centering
    \includegraphics[width=1\linewidth]{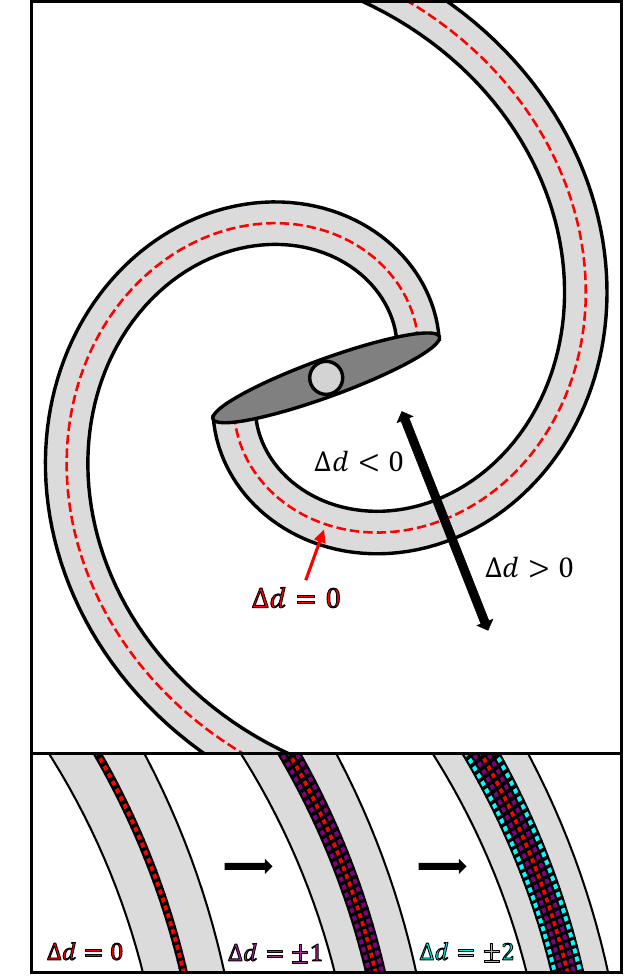}
    \caption{Top: Illustration of the spiral arm spine (red), defined as the geometrical central ridge of the spiral arm masks. We assign negative distances to the spiral arm spine towards the centre of the galaxy, and positive values for the outer side, corresponding to the upstream and downstream side for NGC\,4321 and M51.
    Bottom: Illustration of how spiral arm spine offsets $\Delta d$ are measured in practice using dilation.}
    \label{fig:method}
\end{figure}

\subsection{Spiral arm spine offset and phase bin segmentation}

As illustrated in Fig.~\ref{fig:method}, we first extracted the spines of the spiral arms (the geometrical central ridge of the spiral arm masks) by computing first their medial axis (using \texttt{morphology.medial\_axis} from \texttt{skimage}) and then their skeleton (using \texttt{FilFinder2D.skeleton\_longpath} from \texttt{fil\_finder}). 
    Starting from a binary mask representing the spine, successive morphological dilations generate concentric layers around it. By iteratively applying dilation (\texttt{morphology.dilation} from \texttt{skimage}) and accumulating the dilation steps, each pixel was assigned a value corresponding to its projected distance, $\Delta d$, from the respective spine (which corresponds to $\Delta d=0$\,pc). Hence, $\Delta d$ defines the distance to the spiral arm spine, which is similar to the underlying piecewise logarithmic function constructed by \citet{Querejeta_2021}. We assign negative $\Delta d$ values to the inner side of the spiral arms (i.e. towards the centre of the galaxy) and positive values for the outer side. 
    
    Within the field of views (FOVs) of both NGC\,4321 and M51, the rotational velocity, at $150 -250$\,km/s \citep{Lang_2020} and $200-220$\,km/s \citep{Colombo_2014b}, respectively, exceeds the spiral pattern velocity, at $\sim$ 28\,km/s/kpc \citep{Rand_2004} and $\sim$ 61\,km/s/kpc \citep{Font_2024}. Thus, negative $\Delta d$ values correspond to the upstream and positive values to the downstream side of the spiral arms. We note that this picture might not be so simple for M51, where several Lindblad resonances have been found within the inner few kpc. 
    However, for simplicity, we refer to the inner part of the spiral arms as upstream and the outer part as downstream for both galaxies. For a more detailed analysis of the spiral arm dynamics in M51, we refer to \citet{Sakhibov_2025} and references therein.
    
    The end parts of the spiral arms were cut perpendicular to the spine to retain the full spiral arm width across the entire length of the spiral arms. We applied the spine offset maps in Sects.~\ref{sect:linguine_4321} and \ref{sect:linguine_51}. In addition, we segmented the spiral arms into equally spaced azimuthal bins, hereafter referred to as phase bins, in accordance with the definition of the phase of logarithmic spirals along the spiral arms. This allowed us to investigate whether 
    the observed trends are driven by local variations within the spiral arms or whether they are instead consistent along the entire length of the spiral arms (and thus a feature driven by large-scale spiral arm dynamics). This investigation is described in Sects.~\ref{sect:ratio_4321} and \ref{sect:ratio_51} .
    We split the arms into phase bins of $\Delta\phi \sim \SI{2.5}{\kilo\parsec}$ for NGC\,4321 and of $\Delta\phi \sim \SI{1}{\kilo\parsec}$ for M51 along the spiral to obtain a comparable number of phase bins for both galaxies. 
    The number of phase bins was chosen to be large enough to average over sufficiently many independent sight lines, but small enough to have at least six bins per spiral arm. 

    While constructing the spiral arm masks, \citet{Querejeta_2021} found that the ridges of molecular gas can be aptly fitted by log-spiral functions in NGC\,4321. However, these molecular gas ridges show substantial kinks in M51, potentially caused by the interaction with NGC\,5195, which yields substantial local deviations of the molecular gas ridge from the idealised log-spiral function. 
    Therefore, being based on the log-spiral function, the spine used in this work did not perfectly follow the gas ridge. To account for these deviations, we employed phase bins to mitigate averaging trends across these kinks. 
    
\begin{figure}
    \centering
    \includegraphics[width=1\linewidth]{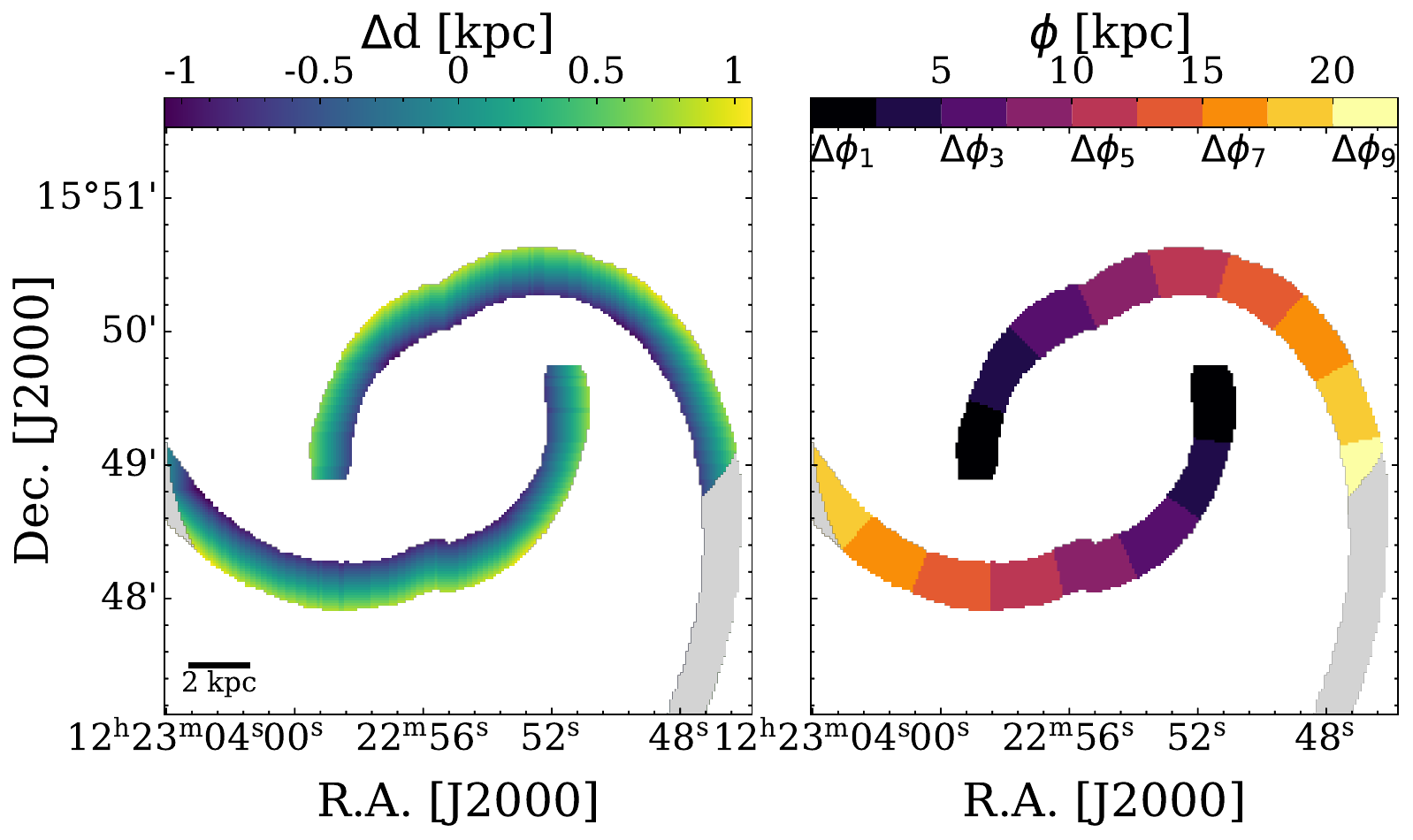}
    \caption{ $\Delta d$ map (left) and phase bins (right) for NGC\,4321 used for the binning in this work. The phase bins have sizes of $\Delta\phi = \SI{2.5}{\kilo\parsec}$ along the spine. Their colour scheme is applied in Fig.~\ref{fig:NGC4321_ratio_delta_d_phase}. The grey segments lie beyond the FOV of the HCN observations.}
    \label{fig:NGC4321_deltad_phase}
\end{figure}

\begin{figure}
    \centering
    \includegraphics[width=1\linewidth]{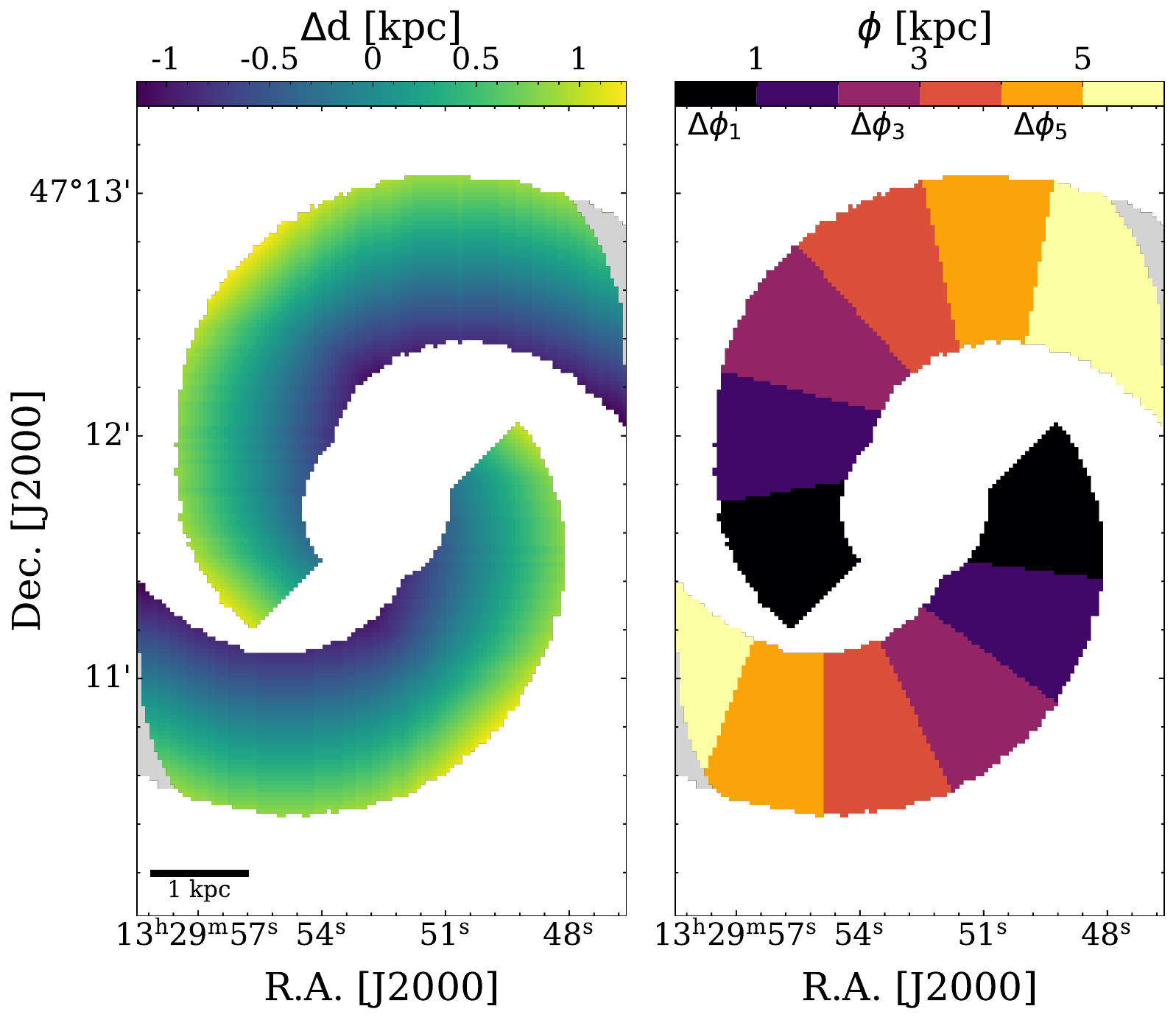}
    \caption{Same as Fig.~\ref{fig:NGC4321_deltad_phase}, but for M51.
    The phase bins have sizes of $\Delta\phi = \SI{1}{\kilo\parsec}$ along the spine. Their colour scheme is applied in Fig.~\ref{fig:M51_ratio_delta_d_phase}. The grey segments lie beyond the FOV of the HCN observations.}
    \label{fig:M51_deltad_phase}
\end{figure}

    To ensure that our results were not affected by the influence of the active galactic nucleus (AGN) in the centre of M51 \citep{Stuber_2023}, we excluded the data at galactocentric radii of $R_\text{gal}<\SI{800}{\parsec}$ \citep{Querejeta_2016b, Querejeta_2019}. The resulting spine offset and phase bins segmentation maps are shown in Fig.~\ref{fig:NGC4321_deltad_phase} and \ref{fig:M51_deltad_phase} for NGC\,4321 and M51, respectively.

\section{Results}\label{sect:results} 
To study the average trends of the integrated intensities, $W$ and SFR, we considered pixels where HCN has been detected with a signal-to-noise ratio (S/N) $\geq 3$. For these 3$\sigma$ selected pixels in each $\Delta d$ range, we computed the median, either along the entire spiral arm length (see Sect.~\ref{sect:linguine_4321}, \ref{sect:linguine_51}) or separately for each spiral arm phase segment (see Sect.~\ref{sect:ratio_4321}, \ref{sect:ratio_51}).

This strict S/N threshold, based on the least significant tracer (i.e. HCN), was applied to all maps. This enabled us to compare the trends for CO, HCN, and SFR seen across regions where dense molecular gas is present. For the ratios (i.e. HCN/CO, SFR/CO, SFR/HCN), we included pixels where HCN has S/N $\geq 1$. We then computed the median across these pixels. This looser S/N threshold (compared to the intensity trends above) allows for the inclusion of a much larger fraction of pixels within the spiral arms, which gives a less biased trend compared to clipping most of the data.
We note that for both S/N thresholds, the enforced S/N clipping biases our results towards regions of bright HCN. Nevertheless, we adopted S/N-clipped trends, rather than using the full pixel set, because line ratios can show large excursions when low S/N pixels are included. A comparison with the unbiased but low S/N average trends is presented in Figs.~\ref{fig:NGC4321_unbiased} and~\ref{fig:M51_unbiased} for NGC\,4321 and M51, respectively, which supports our conclusions.

\subsection{NGC\,4321}
\subsubsection{Linguine plots for NGC\,4321}\label{sect:linguine_4321} 
\begin{figure*}
    \centering
    \includegraphics[width=\linewidth]{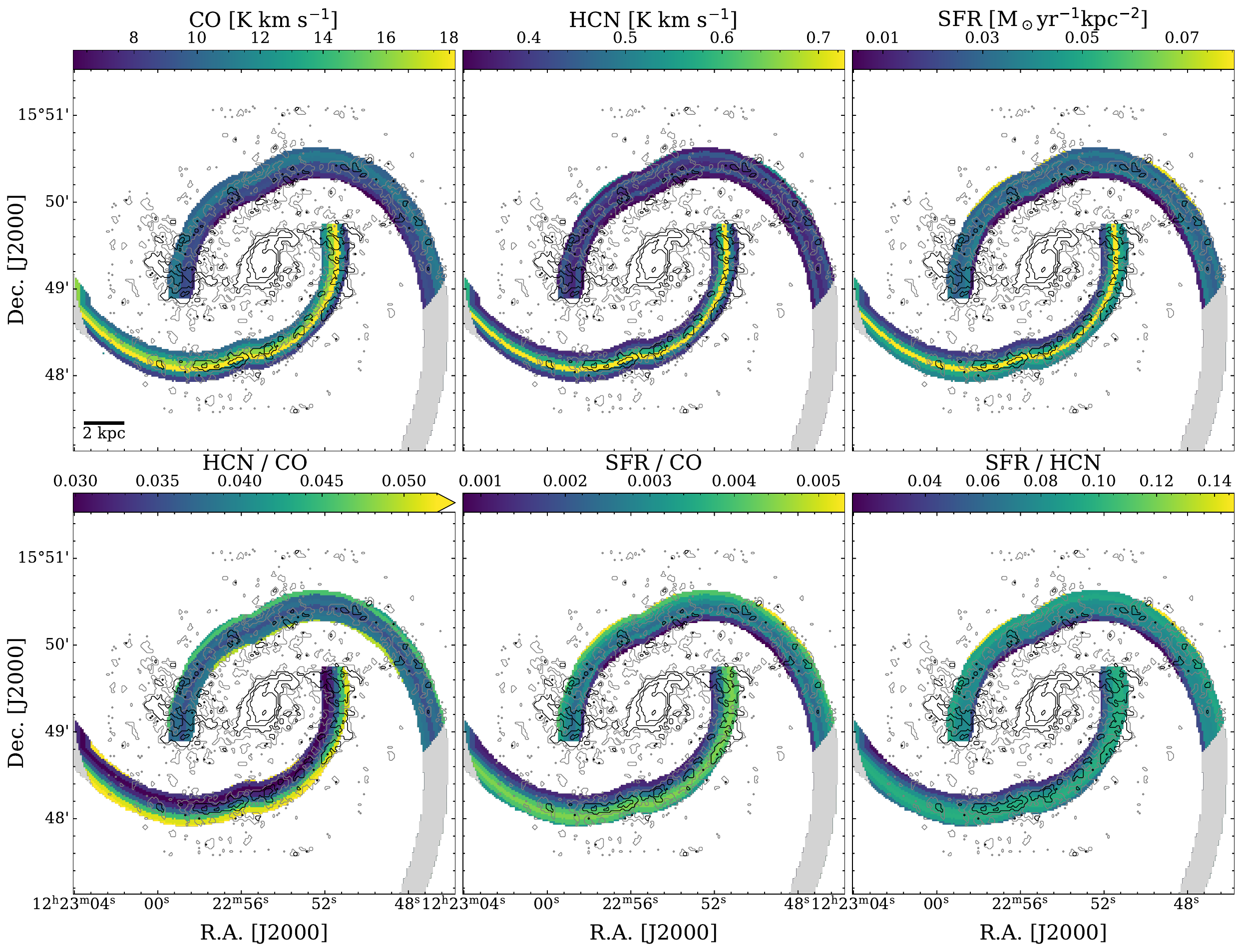}
    \caption{Linguine plot NGC\,4321: Maps of NGC\,4321 displaying the medians binned over $\Delta d$-intervals of $\SI{270}{\parsec}$ in accordance with the HCN beam size computed along the entire spiral arm length. Top row: Distributions of the median integrated intensities of CO, HCN and SFR (from left to right), masked with a S/N = 3 cut based on HCN. Bottom row: Ratios HCN/CO, SFR/CO, and SFR/HCN, based on a S/N = 1 masking based on HCN. The black contours show an HCN S/N of at least 3, 10, and 30. The grey contours additionally show an HCN S/N of at least 1.}
    \label{fig:Linguine_NGC4321}
\end{figure*}

Figure~\ref{fig:Linguine_NGC4321} displays maps of the median integrated intensities of CO, HCN, SFR, and their ratios computed along the entire spiral arm length of NGC\,4321. We refer to this visualization technique as Linguine plots because the resulting structures resemble strands of linguine. The integrated intensities of CO and HCN are much brighter (on average by a factor of 1.4 and 1.3, respectively) in the southern, as compared to the northern spiral arm. In the southern spiral arm, we observe that both CO and HCN peak in the centre of the spiral arm and decrease away from the centre spine (this means with increasing $|\Delta d|$). The increased accumulation of gas at the spine can be explained by the deep potential well due to stars that can be found in the centre of the spiral arm. This leads to more bulk molecular gas (CO) as well as more dense gas (HCN). In the northern arm, we find a slight positive gradient of CO with increasing $\Delta d$. 
We find that SFR (upper right) peaks at the spine in the southern arm, but but shows no corresponding peak in the northern spiral arm. Instead, at high radius, there is an excess of SFR at the outer edge of the arm, driving this trend. This will be discussed in more detail in Sect.~\ref{sect:ratio_4321}.

To study the processes within the spiral arms further, we investigate the trend of the ratios of HCN, CO, and SFR across the spiral arms in the second row of Fig.~\ref{fig:Linguine_NGC4321}. As indicated by \citet{Neumann_2023a}, we find a gradient in the dense gas fraction tracer HCN/CO\,$=(0.030-0.053)$ across the southern spiral arm. However, this is not reflected in the northern arm, which shows a flat trend. We discuss possible causes for a discrepancy between spiral arms in Sect.~\ref{sect:discussion}. The figure also shows that both SFE$_\text{mol}$ and SFE$_\text{dense}$ tend to increase as we move outwards perpendicular to the arm spine (with increasing $\Delta d$). The trend is more pronounced in the northern spiral arm than in the southern, where it flattens just behind the spine. This will be further discussed in Sect.~\ref{sect:ratio_4321}.

\subsubsection{Ratio trends in NGC\,4321}\label{sect:ratio_4321}
\begin{figure*}
    \centering
    \includegraphics[width=1\linewidth]{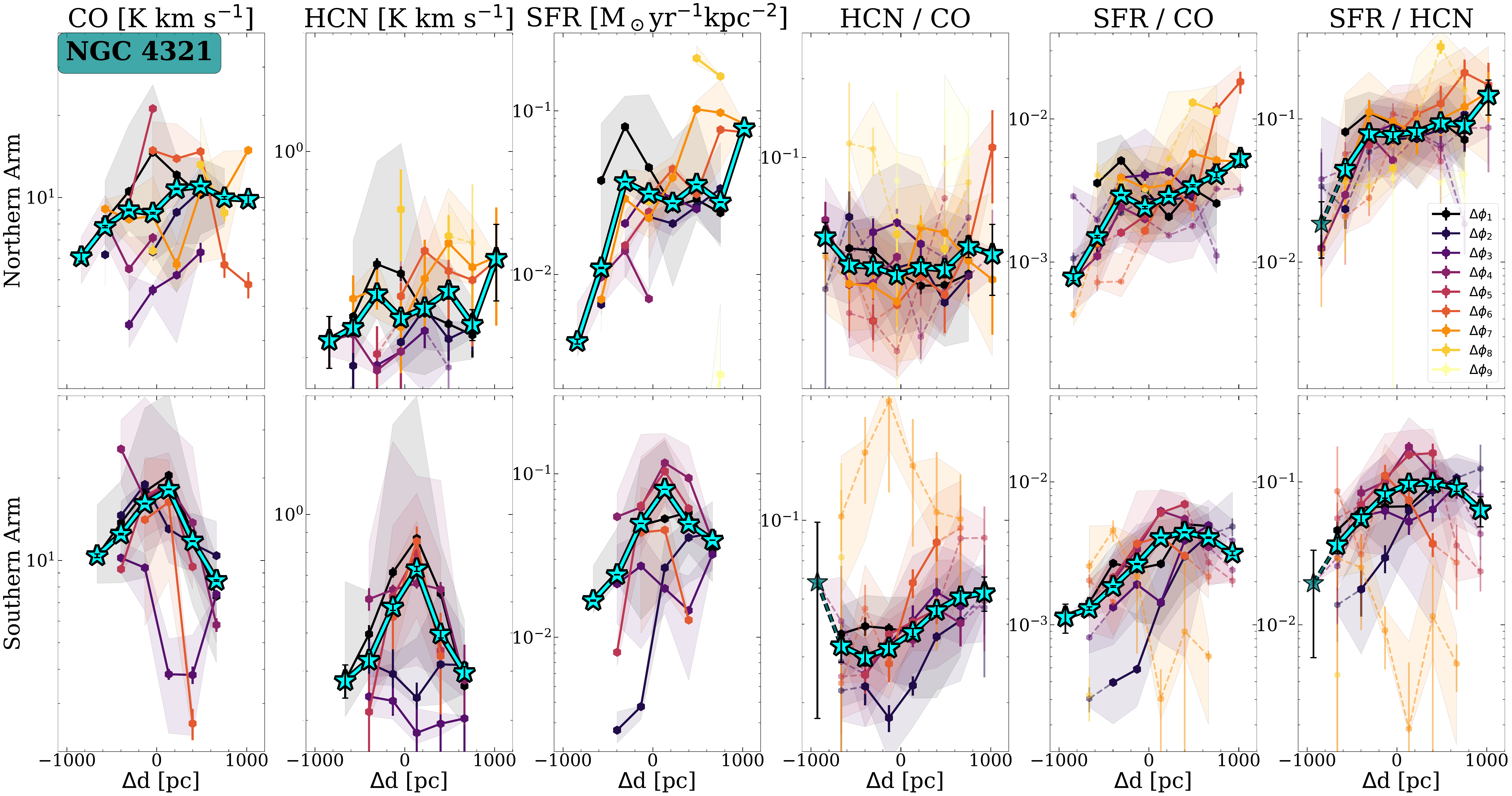}
    \caption{Left three columns: Median trends of the integrated intensities of CO, HCN, and the SFR as a function of the distance to the spiral arm spine ($\Delta d$) in NGC\,4321. Right three columns: HCN/CO, SFR/CO, and SFR/HCN.
    Top (bottom) row: Trends for the northern (southern) spiral arm. The median trends (cyan stars) are computed from data with HCN $\mathrm{S/N}\geq 3$ for CO, HCN, SFR, and HCN $\mathrm{S/N}\geq 1$ for the ratios: HCN/CO, SFR/CO, and SFR/HCN.
    The additional coloured lines show the median trends over $\SI{2.5}{\kilo\parsec}$-wide phase bins using the colour scheme introduced in Fig.~\ref{fig:NGC4321_deltad_phase}, where the shaded areas indicate the 1$\sigma$ scatter of the data in the respective bin. Dashed lines and smaller, fainter markers are used for medians with  S/N$<3$.}
    \label{fig:NGC4321_ratio_delta_d_phase}
\end{figure*}

Spiral arms are regions of deep gravitational potentials across the discs of galaxies. Hence, we expect an accumulation of gas as well as an increase in star formation towards the centre of spiral arms, though with some spatial offset between the potential minimum, traced by $\Sigma_\star$, and the shock-induced peaks of gas (CO and HCN) and star formation (H$\alpha$; e.g. \citealt{Roberts_1969}).

In the median trends computed over the full spiral arm length (cyan stars), we find that CO, HCN, and SFR peak towards the centre of the southern spiral arm ($\Delta d\approx 0$\,pc), as seen in Fig.~\ref{fig:NGC4321_ratio_delta_d_phase}. In contrast, these quantities show a much flatter trend in the northern spiral arm, where we observe a mild increase in CO, HCN (factor of $\approx 2$) and a significant rise in SFR (factor of $\approx 10$) with increasing $\Delta d$ (i.e. inside-out). The flatter behaviour within the northern spiral arm could be due to a less deep gravitational potential compared to the southern spiral arm. This is supported by the stellar mass surface density ($\Sigma_*$) having a more pronounced peak at $\Delta d\approx0$\,pc in the southern arm than in the northern arm (see Fig.~\ref{fig:NGC4321_hcop_Mstar} for details).

Figure~\ref{fig:NGC4321_ratio_delta_d_phase} also shows the median trends in the respective spiral arm segments, computed in increments of \SI{2.5}{\kilo\parsec} along the spiral. For illustrative purposes, these trends are presented on maps that resemble the Linguine plots in Fig.~\ref{fig:Rigatoni_NGC4321}.
Most of the segments agree well with the overall trend across the spiral arms, though with larger scatter.
However, we find significant deviations for two bins in the southern spiral arm ($\Delta\phi_{2,3}= 2.5 -\SI{7.5}{\kilo\parsec}$), where the HCN intensity does not peak towards the centre of the spiral arm. Instead, it reaches a minimum: for at least one bin, the upper edge of the $1\sigma$ shaded area is lower than the lower edge of the $1\sigma$ shaded region for all other phase bins. The CO intensity shows comparable behaviour in $\Delta\phi_{3}$. 

We note that HCN is brighter in the outer edge of the spiral arms ($\Delta d>\SI{500}{\parsec}$) and, hence, more frequently detected (five of nine bins) compared with the inner edge of the spiral arms ($\Delta d<\SI{500}{\parsec}$), where HCN is only detected in three bins. The trend of SFR across the northern spiral arm shows evidence of a second peak at large radii, which yield a higher average SFR  for the outer bins at $\Delta d\geq500$\,pc. These data points originate from the outer segments of the northern arm ($\Delta\phi_{6-8}\geq12.5$\,kpc), whereas no detections are present in these outer phase bins of the southern arm.

Due to the width of the spiral arm masks not being perfectly constant along the full spiral arm length, 
not all phase bins in the northern arm are covered by the mask at $\Delta d\approx1\,$kpc. Thus, the  lower number of pixels, which are contributing to this bin, reduces the significance of the median, as can be seen by the relatively large error bars in Fig.~\ref{fig:NGC4321_ratio_delta_d_phase}. However, in each case (of CO, HCN, and SFR) the median succeeds a S/N of 3 at $\Delta d\approx1\,$kpc and the increased SFR values are consistently found in three phase bins for at least two $\Delta d$-bins each.

The median trends of the ratios (HCN/CO, SFR/CO, SFR/HCN) are shown in the right panels of Fig.~\ref{fig:NGC4321_ratio_delta_d_phase}. The dense gas fraction tracer, HCN/CO, is relatively constant across the northern spiral arm, where it varies by less than a factor of 1.4, while the SFR/CO and SFR/HCN increase with $\Delta d$ (inside-out) by about one order of magnitude. These trends are consistently found across most spiral arm segments, within a scatter of $\sim0.2$\,dex ($\sigma_\text{HCN/CO}=0.18$\,dex, $\sigma_\text{SFR/CO}=0.24$\,dex, $\sigma_\text{SFR/HCN}=0.21$\,dex). 
In the southern arm, we find that all three ratios (HCN/CO, SFR/CO, and SFR/HCN) increase from $\Delta d=\SI{-500}{\parsec}$ to $\SI{500}{\parsec}$, consistent over most spiral arm segments, except for the outermost segment at $\Delta \phi_7 =(15-17.5)$\,kpc. This outermost segment is, on the one hand, affected by low S/N of HCN and, on the other hand, it is close to the corotation radius of the spiral arms of NGC\,4321 at $R_\text{gal}\approx\SI{8}{\kilo\parsec}$ \citep{Elmegreen_1992, Garcia-Burillo_1994}. The implications of this are discussed in Sect.~\ref{sect:discussion}.

Moreover, we note that the increase in HCN/CO is smaller (factor of $\sim 2$), compared to SFR/CO and SFR/HCN (factor of $\sim 4$ and 5, respectively), which points towards a tighter spatial correlation between HCN and CO than between the SFR and the molecular gas tracers. Despite the difference in CO, HCN, and SFR brightness between the two arms, the build-up of dense gas and the SFE transverse to the arms appear to be consistent in both arms.

\subsection{M51}
We applied the same analysis to the interacting galaxy M51, one of the most studied galaxies in the nearby universe. The spiral nature of M51 and the ways it might relate to density wave theory have been thoroughly investigated with CO, making it a great comparison galaxy \citep{Tully_1974, Elmegreen_1989, Vogel_1993, Meidt_2008, Colombo_2014b, Schinnerer_2017}. We note that along with NGC\,4321, these are the only two grand-design spirals beyond the local group with dense gas tracer observations at $\lesssim \SI{300}{\parsec}$ resolution across a significant area of the molecular gas disc. The key differences between NGC\,4321 and M51 are that NGC\,4321 is barred and non-active, while M51 has an AGN, no bar, and its spiral arms are tidally induced by its companion NGC\,5195.

\subsubsection{Linguine plots for M51}\label{sect:linguine_51}
\begin{figure*}
    \centering
    \includegraphics[width=0.7\linewidth]{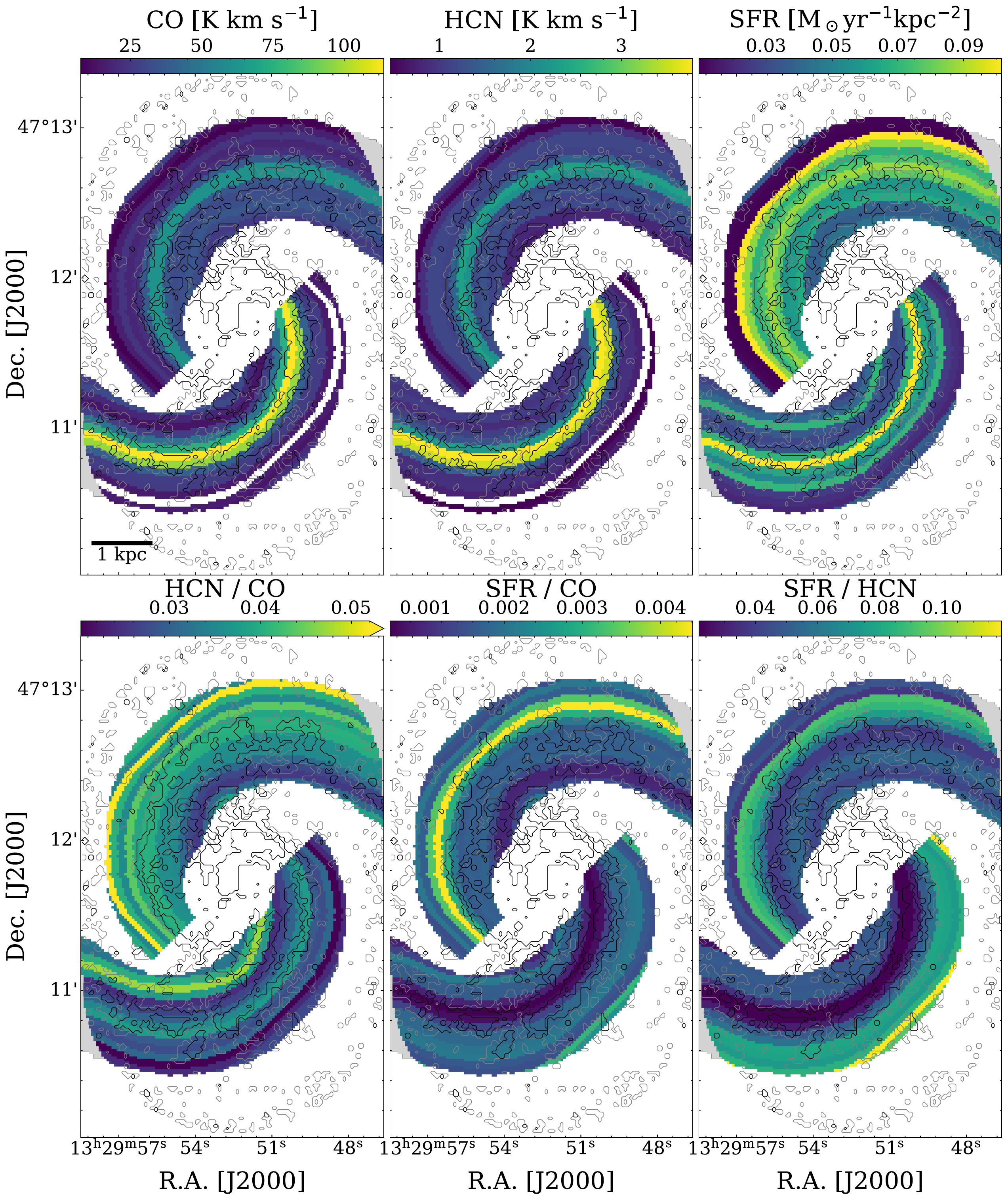}
    \caption{Linguine plot M51: Maps of M51 displaying the medians binned over $\Delta d$-intervals of $\SI{125}{\parsec}$ in accordance with the HCN beam size computed along the entire spiral arm length. Top row: Distributions of CO, HCN, and SFR, masked with a S/N = 3 cut based on HCN. Bottom row: Ratios HCN/CO, SFR/CO, and SFR/HCN, based on a S/N = 1 masking based on HCN. The black contours indicate an HCN S/N of at least 3, 10 and 30, while the grey contours additionally show regions with S/N of at least 1.}
    \label{fig:Linguine_M51}
\end{figure*}
Figure~\ref{fig:Linguine_M51} (Linguine plots) presents maps of the median variation of HCN, CO, and SFR (top panels) and HCN/CO, SFR/CO, and SFR/HCN (bottom panels) similarly to NGC\,4321, adopting the methodology described in Sect.~\ref{sect:method}.

In M51, the northern spiral arm shows a relatively flat trend in CO and HCN line intensity, peaking close to the centre of the spiral arm. These maxima in CO and HCN are more pronounced in the southern than in the northern spiral arm. The SFR shows a clear gradient across the northern arm, increasing inside-out and peaking far behind the spine. This results in small gradients of SFR/CO and SFR/HCN, which both increase inside-out. In the southern arm, the SFR reaches its maximum approximately at the spiral arm spine, where we found the lowest SFR/CO and SFR/HCN, while larger values were seen on the outer and inner edge.

\subsubsection{Ratio trends in M51}\label{sect:ratio_51}
\begin{figure*}
    \centering
    \includegraphics[width=1\linewidth]{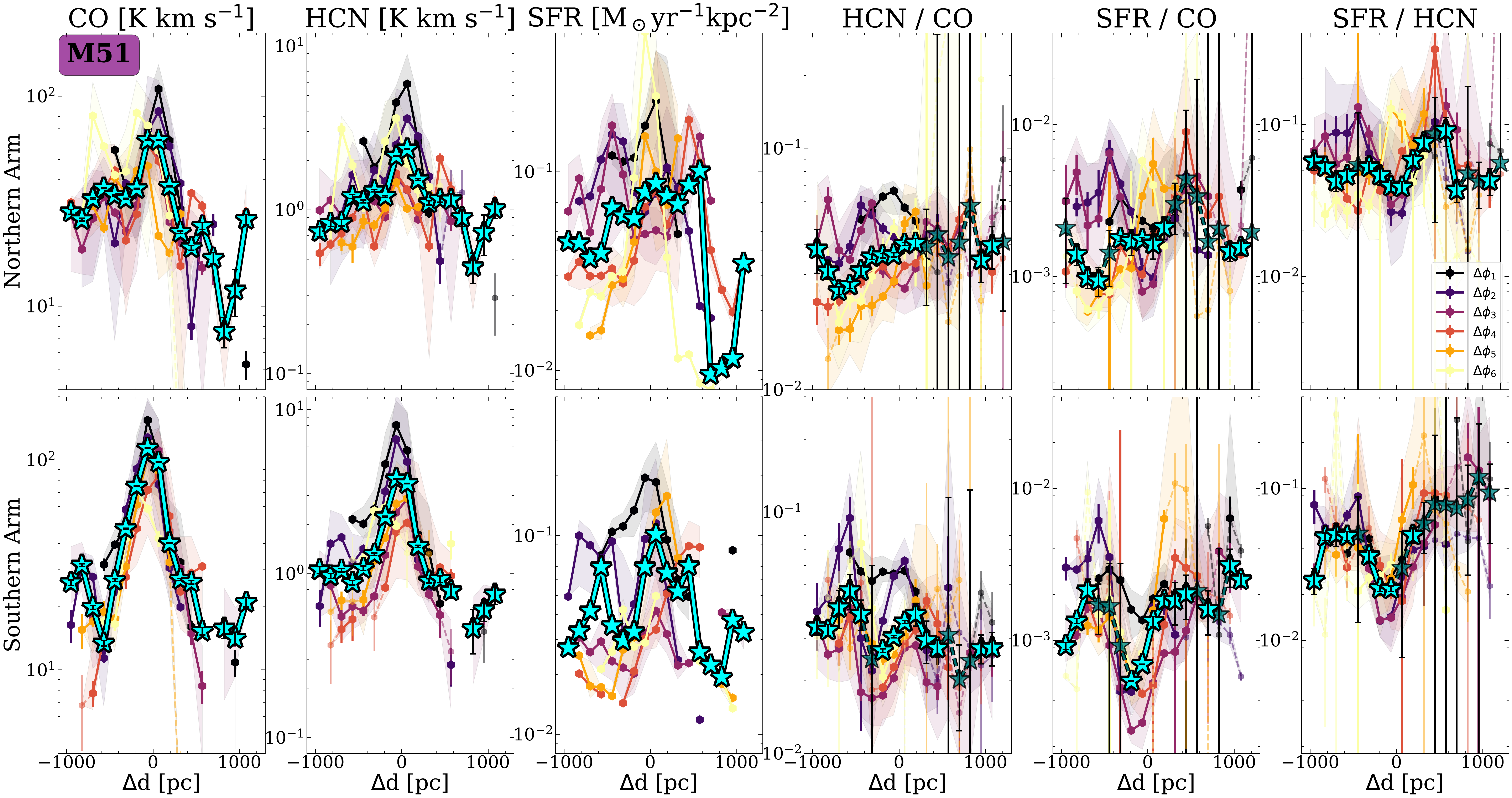}
    \caption{M51 medians of CO, HCN, SFR and HCN/CO, SFR/CO, SFR/HCN binned over $\Delta d$ in cyan stars. Top (bottom) row: Trends for the northern (southern) spiral arm. The median trends are computed from data with HCN $\mathrm{S/N}\geq 3$ for the CO, HCN, SFR, and HCN $\mathrm{S/N}\geq 1$ for the ratios: HCN/CO, SFR/CO, and SFR/HCN. 
    The additional coloured lines show the median trends over $\SI{1}{\kilo\parsec}$-wide phase bins using the colour scheme introduced in Fig.~\ref{fig:M51_deltad_phase}, where the shaded areas indicate the 1$\sigma$ scatter of the data in the respective bin. Significant medians (S/N $\geq 3$) are shown with a solid line. Bins with a low significance are displayed with faint markers and dashed line.}
    \label{fig:M51_ratio_delta_d_phase}
\end{figure*}

In Fig.~\ref{fig:M51_ratio_delta_d_phase}, we can see that the trend in the CO and HCN line intensities in the individual spiral arm segments is in very good agreement with the median trend across both spiral arms. The innermost phase bins $\Delta\phi_{1,2}$ show the most pronounced peaks at the spine (at $\Delta d \approx \SI{0}{\parsec}$). For visualization, the trends in the phase bins are shown on maps similar to the Linguine plots in Fig.~\ref{fig:Rigatoni_M51}.
The maximum intensities of CO and HCN are brighter by factors of 1.8 and 1.6, respectively, in the southern compared to the northern spiral arm. As mentioned in Sect.~\ref{sect:ratio_4321}, this could be caused by a deeper gravitational potential of the southern spiral arm. As shown in Fig.~\ref{fig:M51_hcop_Mstar}, $\Sigma_*$ remains constant across the northern spiral arm for $\Delta d\lesssim0$\,pc before it decreases at $\Delta d\gtrsim 100$\,pc. The southern arm displays a more pronounced trend, reaching its peak shortly after the spine. The implications of this are further discussed in Sect.~\ref{sect:discussion}.

In general, we did not find stringent, monotonic trends for HCN/CO, SFR/CO, and SFR/HCN in the spiral arms of M51 (Fig.~\ref{fig:M51_ratio_delta_d_phase}). The dense gas fraction tracer, HCN/CO, tends to increase in the northern spiral arm with increasing $\Delta d$ and shows the opposite trend in the southern spiral arms; however, it still remains well within a factor of 2.5, with a scatter of 0.22\,dex.
The star formation efficiency tracers, SFR/CO and SFR/HCN, display no strong trend on average in the northern arm; whereas in the spiral arm segments SFR/CO and SFR/HCN vary within a scatter of 0.37\,dex and 0.34\,dex, respectively, around median values of 0.0021 and 0.052 M$_\odot$\,yr$^{-1}$\,kpc$^{-2}$\,/\,K\,km\,s$^{-1}$. These flat trends at $\Delta d\lesssim0$\,pc can be traced back to the opposing behaviour of the outermost phase bins $\Delta\phi_{5,6}$, showing an increase at $\Delta d\lesssim0$\,pc, while the other segments drop there. This difference could potentially be caused by the kinks in the gas ridge mentioned in Sect.~\ref{sect:method}.

In the southern spiral arm, we find a dip of SFR/CO and SFR/HCN just before the centre of the spiral arm (at $\Delta d \approx \SI{-250}{\parsec}$). Increasing outwards, this means with increasing $\Delta d$, we find an average increase in SFR/CO and SFR/HCN by a factor of $\approx 5$. Overall, the individual phase bins along the spiral support the average trends across the full spiral arm, in particular, the dip and subsequent increase in SFR/CO and SFR/HCN is also present in all of the individual segments, showing that the average trends are not driven by a small section of the spiral arms. 

However, we note that the observations of M51 cover a much smaller physical area than those of NGC\,4321 ($R_\text{gal}\lesssim 0.34 R_{25}$ for M51 and $R_\text{gal}\lesssim 0.87 R_{25}$ for NGC\,4321 within the spiral masks) and, hence, they only cover spiral arms lengths of about \SI{7}{\kilo\parsec}. Due to its strong bar, NGC\,4321's spiral arms start at larger radii compared to those of M51, so that the spiral spiral masks in M51 cover a much large fraction of the FOV in M51 than in NGC4321. We compare the trends across the spiral arms with the radial trend in Sect.~\ref{sect:radial}, and we discuss the galaxy comparison further in Sect.~\ref{sect:comparison}. 

\begin{figure*}
    \centering
    \includegraphics[width=1\linewidth]{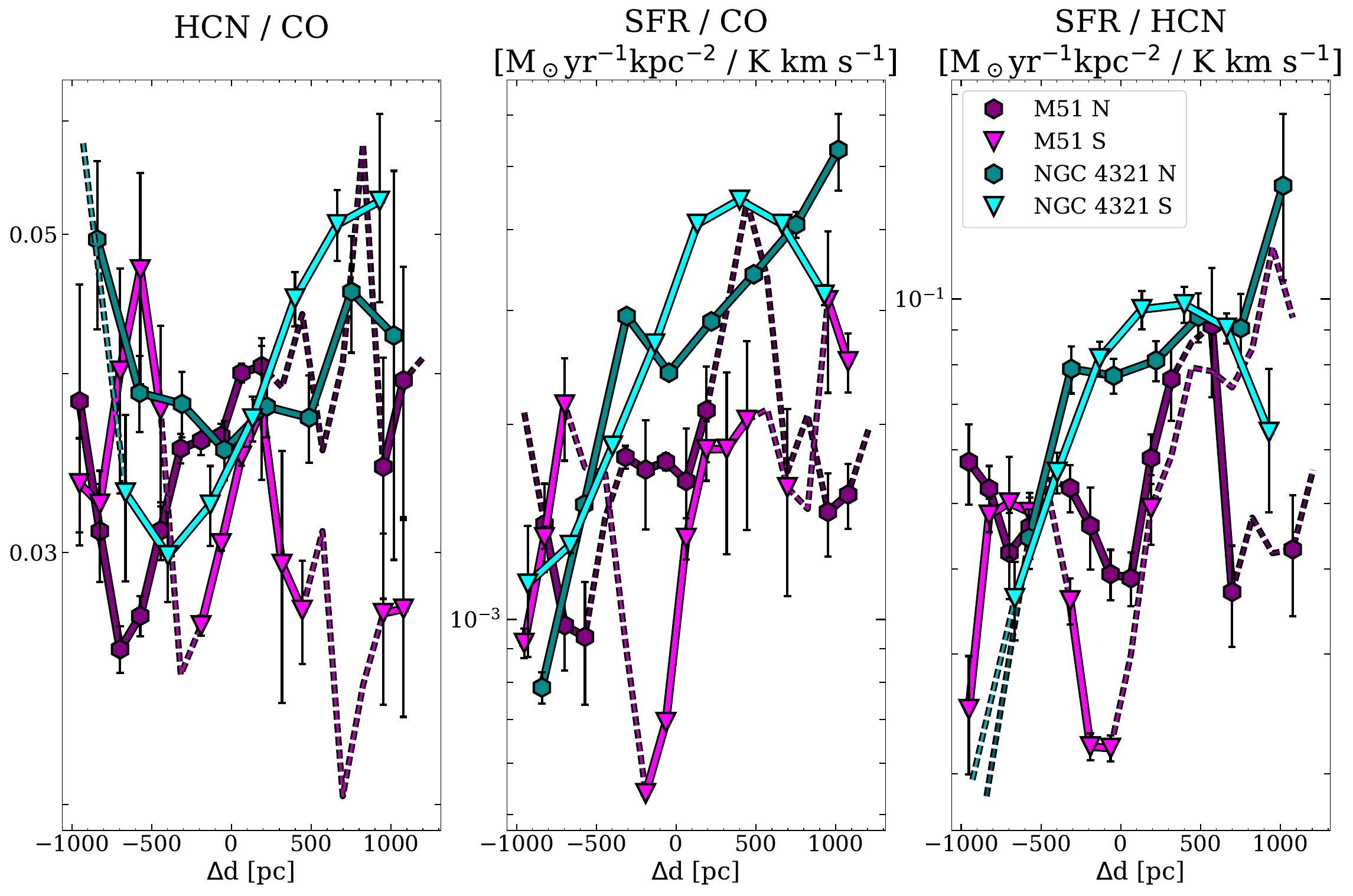}
    \caption{Comparing the medians binned over $\Delta d$ of both galaxies for the three fractions tracing $f_\text{dense}$, SFE$_\text{mol}$, and SFE$_\text{dense}$ (from left to right). 
    Each line represents one spiral arm, the galaxy NGC\,4321 is shown in turquoise and M51 in purple. 
    The medians are binned over a $\Delta d$-range in agreement with the galaxy's beam size and take only pixels into account, where HCN succeeds a S/N of 1. Bins with a S/N$>3$ are indicated with a solid line and markers, others with a dashed line. The northern spiral arms are indicated with hexagonal markers and the southern spiral arms with triangles.}
    \label{fig:4arms_4lines}
\end{figure*}

\section{Discussion}\label{sect:discussion}
\subsection{Comparison of NGC\,4321 and M51}\label{sect:comparison}
After analysing the galaxies separately in great detail, we now take a step back and return to our initial science question: determining how spiral arms affect the transformation from molecular to dense gas and subsequently into stars, and whether there are gradients of the SFEs across the spiral arms.

To investigate this further, we show the median trends of HCN/CO, SFR/CO and SFR/HCN for both spiral arms of both galaxies together in Fig.~\ref{fig:4arms_4lines}. 
Overall, the median ratios span a similar range in both galaxies. The dense gas fraction tracer HCN/CO in the southern spiral arm of NGC\,4321 appears comparable to the northern arm of M51, both showing enhanced HCN/CO at the trailing side of spiral arms; however, it decreases rapidly to their local minima at $\Delta d\approx\SI{-500}{\parsec}$, after which HCN/CO gradually increases across the spiral arms towards the downstream side. The trend in the northern arm of NGC\,4321 is less pronounced. Unfortunately, there are only a few significant medians seen for both arms of M51 at $\Delta d>\SI{500}{\parsec}$, preventing a reasonable comparison on the downstream arm side.

The molecular gas star formation efficiency, traced by SFR/CO, increases in both spiral arms of NGC\,4321 and M51 from the upstream to downstream arm parts (middle panel of Fig.~\ref{fig:4arms_4lines}).
However, the trend is steeper in NGC\,4321.
Furthermore, in the southern arm of M51, the SFR/CO trend is offset by $\sim\SI{500}{\parsec}$ compared to the northern arm and to the trends in NGC\,4321.
This could indicate that for M51, either the stellar mass-based spiral arms do not perfectly follow the molecular gas spiral arms or the formation of dense gas and stars is actually delayed in the southern arm of M51. This asymmetry between the north and south of M51, especially in SFE$_\text{dense}$, has already been observed in previous studies \citep[e.g.][]{Querejeta_2019}. Moreover, the interaction of M51 with its companion (i.e. NGC\,5195 is connected with M51's northern arm) can significantly affect the gravitational potential and cause differences in the dynamics of the spiral arms.

Similarly to the molecular gas SFE, the dense gas star formation efficiency, traced via SFR/HCN, shows an increasing behaviour across the spiral arms (right panel of Fig.~\ref{fig:4arms_4lines}), where M51 once again shows an offset of $\SI{500}{\parsec}$ compared to NGC\,4321. Remarkably, the peak SFR/HCN at the downstream side of the spiral arms is very comparable across the four spiral arms.

In general, we find a more complex picture in M51, where SFR/CO and SFR/HCN do not vary in a stringent manner across the spiral arms. One explanation could be that in M51, we cover a much smaller FOV and, hence, spiral arm length and miss parts of the logarithmically expanding spiral arms.
A wider FOV would be needed to investigate whether the spiral arm segments outside of the central 5\,kpc region show a different behaviour. However, even in the central phase bins of NGC\,4321, we found a behaviour that is distinct to that of M51. 
Despite both galaxies being classified as grand-design spirals \citep{Buta_2015}, the large-scale bar in NGC\,4321 could be driving the spiral structure, as opposed to the small bar in M51. 
However, the presence of the bar further prevents us from limiting the FOV of NGC\,4321 to the same radial interval as in M51 and, thus, we were not able to test the results at a comparable region sampling.
Observations of additional galaxies, both with and without a bar, would be needed to draw firm conclusions.
We note that the spiral arm mask of M51 contains the molecular ring at $R_\text{gal}<1.3$\,kpc, which is expected to behave dynamically different. This could affect the two phase bin segments $\Delta\phi_{1,2}$ near the centre. 

Another reason for the complex behaviour in M51 could be the definition of the spiral arms. The spiral arm mask of M51 has been created based on NIR observations of a much larger FOV (Querejeta et al.\ in prep) and it is likely that it is not perfectly matched to the spiral arm structure in the SWAN FOV studied here. 
Therefore, we find that the spiral arm mask: (a) might not perfectly follow the stellar and gas density structure and (b) might overestimate the spiral arm width in the inner region of the galaxy.
The overestimated spiral arm width might create a misclassification of some material (especially SF) from the other spiral arm in the upstream side of the mask, particularly in the segments nearest to the centre.
Removing this region ($-1000\,\mathrm{pc}<\Delta d<-500\,$pc) from the trends of M51 yields a consistent increase in SFR/CO and SFR/HCN from the upstream to the downstream side, consistent with the trend found in NGC\,4321 if offset by 500\,pc.

\subsection{Radial trend in NGC\,4321}\label{sect:radial}
We found strong positive gradients of approximately one order of magnitude in both SFR/CO and SFR/HCN across the spiral arms of NGC\,4321 as seen in Fig.~\ref{fig:NGC4321_ratio_delta_d_phase}. To ensure that these trends are truly related to the spiral arm spine offset, we compared them with the radial trend analysed by \citet{Neumann_2024}.
In the southern arm we observed a gradient of $\mathrm{SFR/HCN} = (0.036 - 0.098)$\,M$_\odot$\,yr$^{-1}$\,kpc$^{-2}/$\,K\,km\,s$^{-1}$ (factor of 3) perpendicular to the spine over a range of $\SI{1.5}{\kilo\parsec}$. This is significantly stronger compared to the radial trend, which shows a gradient of $\mathrm{SFR/HCN} = (0.08 - 0.2)$\,M$_\odot$\,yr$^{-1}$\,kpc$^{-2}/$\,K\,km\,s$^{-1}$ (factor of 2.5) over a range of $\SI{4}{\kilo\parsec}$, as measured from the mean SFR/HCN trend across the two spiral arms \citep{Neumann_2024}.
The behaviour of HCN/CO is more complex to interpret, as the northern arm shows a much flatter trend than the southern arm of NGC\,4321, where we found a positive gradient of $\mathrm{HCN/CO} = 0.030 - 0.053$ over a range of $\sim\SI{1.5}{\kilo\parsec}$. Although combining both arms would lead to a flatter slope, it would still be much stronger than the radial trend. \citet{Neumann_2024} did not find a significant gradient with the radius, but a flat trend with a scatter of $\mathrm{HCN/CO} = 0.02 - 0.04$ over $>\SI{4}{\kilo\parsec}$ instead. One possible explanation for the difference in the arms could be that this type of gradient is only detectable if the gravitational potential is pronounced enough. Regardless of this difference between the northern and southern arm, we conclude that the observed gradients of HCN/CO and SFR/HCN across the spiral arms do not result from the radial trend and that they are directly linked to the spiral arm offsets instead.

\subsection{Density wave theory and spiral arms}
From the quasi-stationary density wave theory, we expect that within spiral arms and from the upstream to the downstream side: first the molecular gas (CO), then dense gas (HCN), and, finally, stars (SFR) are formed \citep[for example][]{Roberts_1969}. This should result in positive trends of the dense gas fraction (traced by HCN/CO) and the molecular (as well as the dense) gas star formation efficiency (traced by SFR/CO and SFR/HCN) across the spiral arms. This is in agreement with our observations for NGC\,4321 and at least partially for M51. If the offset between CO and SFR is larger than that between HCN and SFR, we would expect larger deviations and, thus, a slightly steeper slope for SFE$_\text{mol}$ than for SFE$_\text{dense}$. Although our findings do not confirm this directly, it could be the case that the resolution of $\sim\SI{270}{\parsec}$ is not high enough to resolve the step from molecular to dense gas. \citet{Kim_2022} have shown that the separation from molecular gas-dominated and star formation-dominated regions occurs typically at a 225\,pc scale and that $\leq150$\,pc scale observations are needed to resolve this separation. 

In the picture of quasi-stationary density wave theory, we might expect a change in the trends of the ratios (HCN/CO, SFR/CO, SFR/HCN) with $\Delta d$ leading to an opposite behaviour beyond corotation \citep{Egusa_2009}. However, we would not expect such a sudden change at $R_\text{CR}$ as depicted by the outermost segment (i.e. $\Delta\phi_7$) in the southern arm of NGC\,4321 and, instead, a flattening followed by a progressively steeper slope with the opposite gradient if we could probe far past corotation. 

Although finding support for the density wave theory in our spiral arm analysis (especially in NGC\,4321) there are also trends in HCN/CO, SFR/CO, and SFR/HCN that cannot be explained by this simple picture.
For instance, pure density wave theory cannot explain the dips of SFR/CO, SFR/HCN close to spiral arm centre in M51 or the differences in the trends between the northern and southern spiral arm of M51.
Instead, these trends might be substantially driven by radial variations of CO, HCN, and SFR that are particularly relevant in the inner few kpc probed in M51.

The nature of the grand spiral galaxy M51 has been a topic of discussion for years. According to \citet{Tully_1974, Elmegreen_1989, Vogel_1993, Meidt_2008, Colombo_2014b}, the inner spiral of M51 could have formed via a spiral density wave.
\citet{Colombo_2014b} find, by using a kinematic decomposition of line of sight velocity field of molecular gas, that the inner spiral pattern is consistent with a spiral density wave, but that it is more consistent with a material arm beyond $R_\text{gal}\sim\SI{100}{\arcsecond}\sim\SI{4.2}{\kilo\parsec}$, which aligns with the corotation radius estimated by \citet{Querejeta_2016b}.
However, \citet{Schinnerer_2017} conclude that the density wave nature of M51's spirals alone cannot explain the observed pattern of star formation, particularly in the northern spiral arm segment of M51. They conduct a sophisticated high-resolution, multi-wavelength study of nine gas spurs, combining observations of the ionized, atomic and molecular gas with tracers of recent star formation. They argue that more localized mechanisms besides the spiral density wave are needed for the star formation to occur in spurs.

There are several factors that might contribute differences from the conventional density wave picture.  For example, flows in and along the spiral arms could prolong the collapse of molecular clouds \citep[and thus SF, see][]{Meidt_2013} until after passage outside the spiral environment. The southern arm of M51 is known to have a lower SFE$_\text{dense}$ despite high $f_\text{dense}$ \citep{Querejeta_2019, Stuber_2025b}, perhaps explaining some of the differences found here between the northern and southern arm. In this type of scenario, the location of star-forming sites relative to the gas spurs (and therefore possible offsets between both) could depend on the depth of the spiral arm potential \citep{Schinnerer_2017}. This would imply that investigated offsets (and, hence, the HCN/CO, SFR/CO, and SFR/HCN variations) are sensitive to the detailed shape of the gravitational potential. 

Another presumably important factor is the gas self-gravity, both for its coordination of star formation and for its influence on the strength of the density wave shock \citep[i.e.][]{Elmegreen_2018, Henshaw_2020}.  
The more strongly self-gravitating the gas, the relatively weaker is its response to the underlying stellar spiral density wave leading to weaker and less frequent offset shocks. Self-gravity becomes increasingly important as the gas mass increases in relation to the stellar mass. We traced the strength of gas self-gravity here in terms of $f_\text{gas}\propto$\,CO\,/\,$\Sigma_*$.
Despite NGC\,4321 and M51 showing similar trends in CO, Fig.~\ref{fig:NGC4321_hcop_Mstar} and \ref{fig:M51_hcop_Mstar} reveal their divergent behaviour in CO\,/\,$\Sigma_*$. M51 shows much stronger peaks at $\Delta d =0$\,pc, particularly in the southern spiral arm, indicating that self-gravity is indeed stronger in M51 relative to the underlying density wave spiral. This could explain the differences between M51 and NGC\,4321. While the gas in the spiral arms of NGC\,4321 appears highly sensitive to the density wave, the relatively high gas self-gravity in M51 gives the density wave a weaker organizational influence on the gas in the spiral arms. 
In this case, shocks may still be present and initiate star formation, but their origin would not solely be to the density wave \citep{Meidt_2020}. This could be the reason why in M51's inner spiral arms some of the  gas nearest to the peak ($\Delta d <500$\,pc) exhibits mildly similar behaviour as observed over a wider area in NGC\,4321.

In addition to such differences between galaxies, the depth of the spiral arm potential in contrast to the self-gravity of the gas can vary across the disc of a galaxy, creating lopsided stellar and gas mass distributions. \citet{Colombo_2014b} argued that in M51, this lopsidedness could be caused by the overlap between different spiral modes. Moreover, local factors like feedback can introduce asymmetries in the gas content from arm to arm, which might possibly explain the differences observed between the northern and southern spiral arm of M51.

\section{Conclusions}\label{sect:conclusion}
We investigated how the proxy for the dense gas fraction HCN/CO, along with proxies of the star formation efficiency of molecular gas and dense gas SFR/CO and SFR/HCN, vary within the spiral arms of two nearby, spiral galaxies, NGC\,4321 and M51. We specifically analysed the variation of these quantities perpendicular to the spiral arm spine ($\Delta d$) to test the evolutionary sequence from molecular gas to dense gas and star formation. Our key findings are as follows: 

\begin{enumerate}
    \item In three out of four spiral arms, tracers of molecular gas (CO, HCN), stars (3.6\,$\mu$m) and star formation (H$\alpha$, 24\,$\mu$m) peak close to the spine of the spiral arms, demonstrating that gas densities and star formation activity is very high in the central areas of spiral arms. Only the northern spiral arm of NGC\,4321 shows an increase in CO, HCN, and SFR with increasing $\Delta d$ beyond the spiral arm spine, indicating a build up of gas and stars from the upstream to the downstream side of the spiral arm.
    \item In NGC\,4321, we find that to first order, HCN/CO, SFR/CO, and SFR/HCN increase from the upstream to the downstream side of both spiral arms. This suggests that on the one hand, the 270\,pc resolution is sufficient to resolve the spatial decorrelation between gas and stars; on the other hand, it implies that there is an evolutionary sequence from the build up of molecular over dense gas to the formation of stars. This is in agreement with the expectation from quasi-stationary density wave theory.    
    \item We note that the trends are much stronger for SFR/CO and SFR/HCN than for HCN/CO, which indicates that the SF phase is spatially more separate to both gas density phases, respectively, than the spatial offset between these two gas phases, traced by CO and HCN, respectively.
    \item We emphasise that the trends in SFR/HCN across the spiral arms of NGC\,4321---in the southern arm, a gradient of $(0.036 - 0.098)$\,M$_\odot$\,yr$^{-1}$\,kpc$^{-2}/$\,K\,km\,s$^{-1}$ over a range of 1.5\,kpc---are significantly stronger than the radial trend (gradient of $(0.08-0.2)$\,M$_\odot$\,yr$^{-1}$\,kpc$^{-2}/$\,K\,km\,s$^{-1}$ over a range of 4\,kpc; \citealp{Neumann_2024}). This result indicates that the disc dynamics perpendicular to the spiral arms are the main driver of these trends.
    \item In M51 we find strong peaks of the gas tracers (CO, HCN) at the centre of the spiral arms ($\Delta d=0$\,pc), but less prominent trends of HCN/CO, SFR/CO, and SFR/HCN. There are some consistencies with NGC\,4321 in limited parts of the $\Delta d$ range.
    \item We conclude that the more typical spiral galaxy NGC\,4321 seems to behave in accordance with the density wave theory, while the inner part of M51 exhibits (at least in part) a different behaviour. This could be due to relatively stronger role of the gas self-gravity, making the gas in the spiral arms less susceptible only to the density wave. 
    
\end{enumerate} 

Our findings show that for some galaxies, large-scale galactic dynamics (e.g. density waves) can induce a sequence of gas density and star formation-to-gas density variations perpendicular to the spiral arms.
This sequence contributes to the increased scatter of spectroscopic ratios such as HCN/CO and SFR/HCN at sub-kpc scales, which cannot be explained by changes in the density probability density function of GMCs as laid out in \citet{Gallagher_2018b, Neumann_2023a}. Instead, these trends are signatures of the gas-to-star formation life cycles as a consequence of the large-scale kinematics of disc galaxies. However, in this study, we have found (at least partial) differences between NGC\,4321 and M51. Hence, a comprehensive picture illustrating how large-scale galactic dynamics such as density waves can drive variations in proxies of the dense gas fraction (e.g. HCN/CO) and star formation efficiency (e.g. SFR/HCN) require a larger sample of $\lesssim 300\,$pc scale observations of nearby, spiral galaxies.

\begin{acknowledgements} We would like to thank the anonymous referee for their
insightful comments that helped improve the quality of the paper.
This work was carried out as part of the PHANGS collaboration. This paper makes use of the following ALMA data:\\
ADS/JAO.ALMA\#2015.1.00956.S,\\ 
ADS/JAO.ALMA\#2017.1.00815.S.\\ 
ALMA is a partnership of ESO (representing its member states), NSF (USA), and NINS (Japan), together with NRC (Canada), NSC and ASIAA (Taiwan), and KASI (Republic of Korea), in cooperation with the Republic of Chile. The Joint ALMA Observatory is operated by ESO, AUI/NRAO, and NAOJ. 
This work made use of data from SWAN, the IRAM large program `Surveying the Whirlpool galaxy at Arcseconds with NOEMA'. This work is based on data obtained by PIs E. Schinnerer and F. Bigiel with the IRAM-30m telescope and NOEMA observatory under project ID M19AA. IRAM is supported by INSU/CRS (France), MPG (Germany) and IGN (Spain).

Funded by the Deutsche Forschungsgemeinschaft (DFG, German Research Foundation) – BI 1546/6-1. 
DC, FB, and ZB gratefully acknowledge the Collaborative Research Center 1601 (SFB 1601 sub-project B3) funded by the Deutsche Forschungsgemeinschaft (DFG, German Research Foundation) – 500700252. 
MQ, MRG, and AU acknowledge support from the Spanish grant PID2022-138560NB-I00, funded by MCIN/AEI/10.13039/501100011033/FEDER, EU.
RSK acknowledges financial support from the ERC via Synergy Grant ``ECOGAL'' (project ID 855130) and from the German Excellence Strategy via the Heidelberg Cluster ``STRUCTURES'' (EXC 2181 - 390900948). In addition RSK is grateful for funding from the German Ministry for Economic Affairs and Climate Action in project ``MAINN'' (funding ID 50OO2206), and from DFG and ANR for project ``STARCLUSTERS'' (funding ID KL 1358/22-1). 
HAP acknowledges support from the National Science and Technology Council of Taiwan under grant 113-2112-M-032-014-MY3.
JPe acknowledges support by the French Agence Nationale de la Recherche through the DAOISM grant ANR-21-CE31-0010 and by the Thematic Action
“Physique et Chimie du Milieu Interstellaire” (PCMI) of INSU Programme National “Astro”, with contributions from CNRS Physique \& CNRS Chimie, CEA, and CNES.
ES acknowledges funding from the European Research Council (ERC) under the European Union’s Horizon 2020 research and innovation programme (grant agreement No. 694343).
\end{acknowledgements}

\bibliography{bibfile} 

\begin{appendix}

\section{Dense gas tracer HCO$^+$} 
\begin{figure*}
    \centering
    \includegraphics[width=\linewidth]{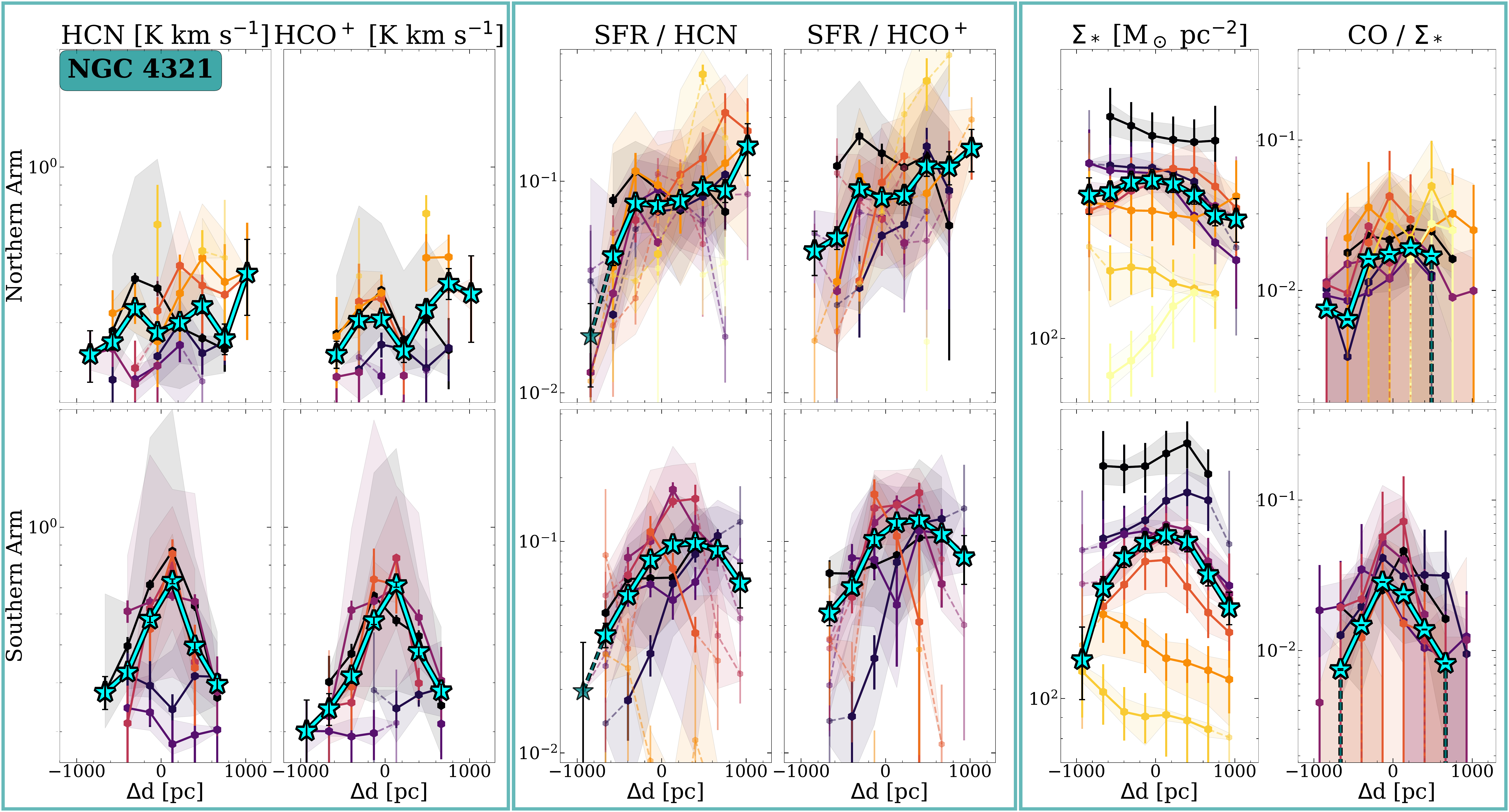}
    \caption{Left column: Median trends of the integrated intensities of HCN and HCO$^+$ as a function of the distance to the spiral arm spine ($\Delta d$) in NGC\,4321. Middle column:  SFR/HCN and SFR/HCO$^+$ in comparison. The median trends (cyan stars) are computed from data with HCN (HCO$^+$) $\mathrm{S/N}\geq 3$ for HCN (HCO$^+$) and HCN (HCO$^+$) $\mathrm{S/N}\geq 1$ for the ratio SFR/HCN (SFR/HCO$^+$).
    Right column: Stellar mass surface density ($\Sigma_*$) tracing the gravitational potential and CO\,/\,$\Sigma_*\sim f_\text{gas}$ tracing the strength of gas self-gravity versus $\Delta d$.
    The additional coloured lines show the median trends over $\SI{2.5}{\kilo\parsec}$ wide phase bins using the colour scheme introduced in Fig.~\ref{fig:NGC4321_deltad_phase}, where the shaded areas indicate the 1$\sigma$ scatter of the data in the respective bin. Dashed lines and smaller, fainter markers are used for medians with  S/N$<3$.}
    \label{fig:NGC4321_hcop_Mstar}
\end{figure*}

\begin{figure*}
    \centering
    \includegraphics[width=\linewidth]{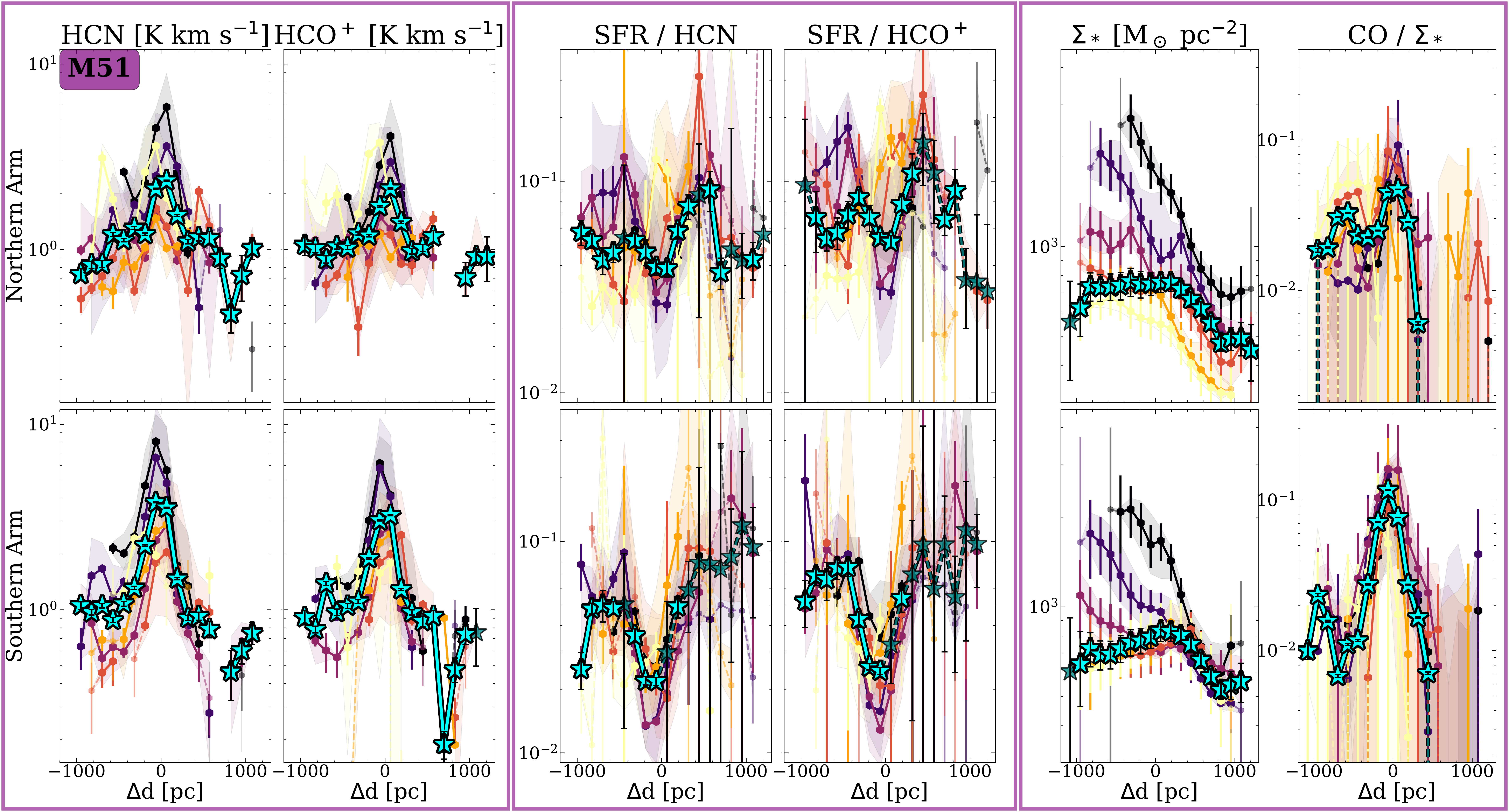}
    \caption{Same as Fig.~\ref{fig:M51_hcop_Mstar} but for M51. The coloured lines show the median trends over $\SI{1}{\kilo\parsec}$ wide phase bins using the colour scheme introduced in Fig.~\ref{fig:M51_deltad_phase}.}
    \label{fig:M51_hcop_Mstar}
\end{figure*}

\cite{Stuber_2025b} recommend HCO$^+$ as a more valid tracer of dense gas, particularly in M51. Therefore we repeat our analysis by replacing HCN with HCO$^+$, which is available for both data sets.
The left panels of Figs.~\ref{fig:NGC4321_hcop_Mstar} and~\ref{fig:M51_hcop_Mstar} show HCN and HCO$^+$, and the middle panels show SFR/HCN in comparison to SFR/HCO$^+$ for NGC\,4321 and M51, respectively. 
We find no significant difference in the trends of HCO$^+$ compared to HCN and SFR/HCO$^+$ compared to SFR/HCN for both galaxies. 

\section{Stellar mass surface density, $\Sigma_*$}\label{sect:Sigma_*}
To investigate trends with the stellar mass surface density ($\Sigma_*$), we consider all available data points, so that we are not biased by the dense gas but instead can study the `smooth' distribution of the stellar mass on galactic scales. In the right panels of Fig.~\ref{fig:NGC4321_hcop_Mstar}, we can see the more pronounced $\Sigma_*$ peak of at $\Delta d\approx0$\,pc, and thus deeper gravitational potential, in the southern arm than in the northern arm of NGC\,4321. From this we would expect stronger deviations in the dense gas, which supports our findings regarding the less pronounced trends in the northern spiral arm of NGC\,4321. 
M51's spiral arms exhibit a similar behaviour, as shown in Fig.~\ref{fig:M51_hcop_Mstar}. In the northern spiral arm, $\Sigma_*$ remains flat for $\Delta d \leq 0$\,pc before it decreases at $\Delta d \geq 0$\,pc, while the southern arm displays a more pronounced trend, where $\Sigma_*$ reaches its peak shortly after the spine.

\section{Impact of S/N clipping}\label{sect:unbiased}
\begin{figure*}
    \centering
    \includegraphics[width=1\linewidth]{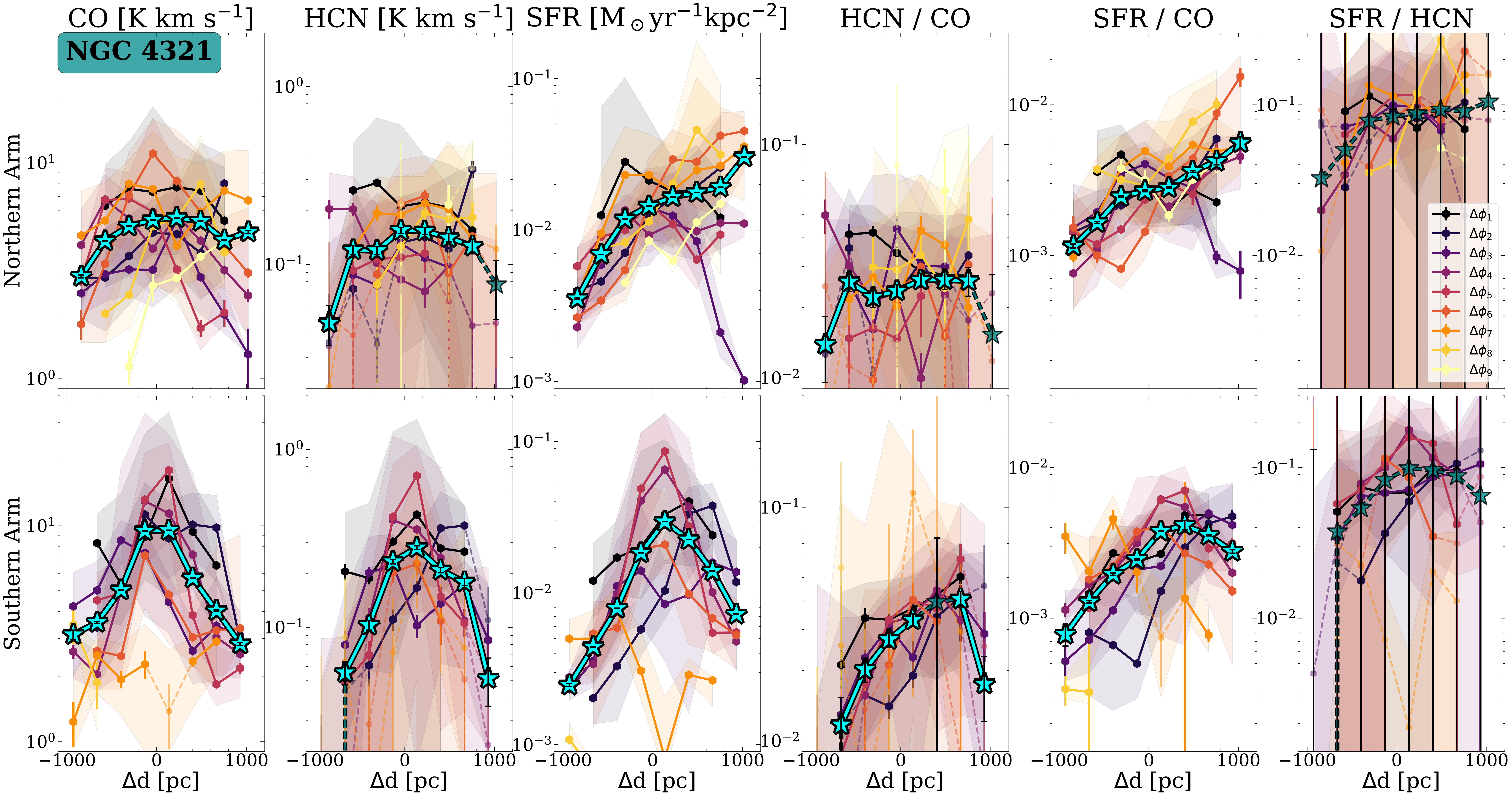}
    \caption{Same data as shown in Fig.~\ref{fig:NGC4321_ratio_delta_d_phase} for NGC\,4321 but presented without S/N clipping to confirm that the results are not significantly affected by bias towards bright HCN regions.}
    \label{fig:NGC4321_unbiased}
\end{figure*}

\begin{figure*}
    \centering
    \includegraphics[width=1\linewidth]{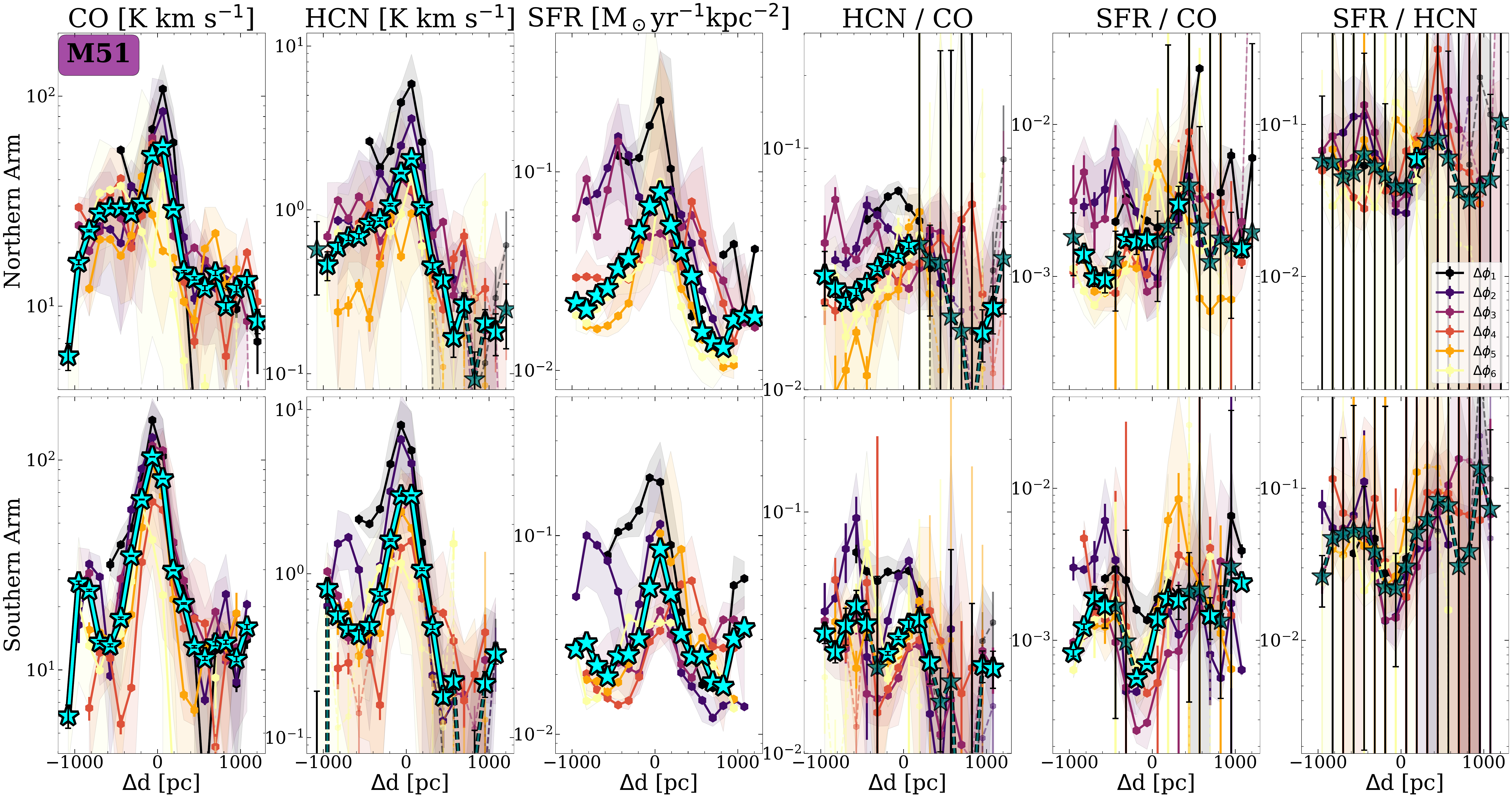}
    \caption{Same data as in Fig.~\ref{fig:M51_ratio_delta_d_phase} for M51 but presented without S/N clipping to confirm that the results are not significantly affected by bias towards bright HCN regions.}
    \label{fig:M51_unbiased}
\end{figure*}

We repeat our analysis without applying any S/N clipping to the data before the binning to investigate how the bias towards bright HCN regions affects our results. Figure~\ref{fig:NGC4321_unbiased} displays the comparison with Fig.~\ref{fig:NGC4321_ratio_delta_d_phase} for NGC\,4321 while Fig.~\ref{fig:M51_unbiased} presents the analogous comparison with Fig.~\ref{fig:M51_ratio_delta_d_phase} for M51.
We find no significant difference in either the average or the phase bin trends for both NGC\,4321 and M51. The difference lies mainly in the lower S/N, which becomes particularly visible in the  trends of SFR/HCN consisting of mostly low-significant bins (S/N$<3$). 
We note that the integrated intensities of CO and HCN remain flat in the northern spiral arm of NGC\,4321 when not biased towards bright HCN regions. However, the previously mentioned additional SFR peak at $\Delta d\approx500$\,pc is even more pronounced without the masking, resulting in a gradient in the average SFR trend across the northern spiral arm of NGC\,4321. 

\section{Rigatoni plots}
\begin{figure*}
    \centering
    \includegraphics[width=1\linewidth]{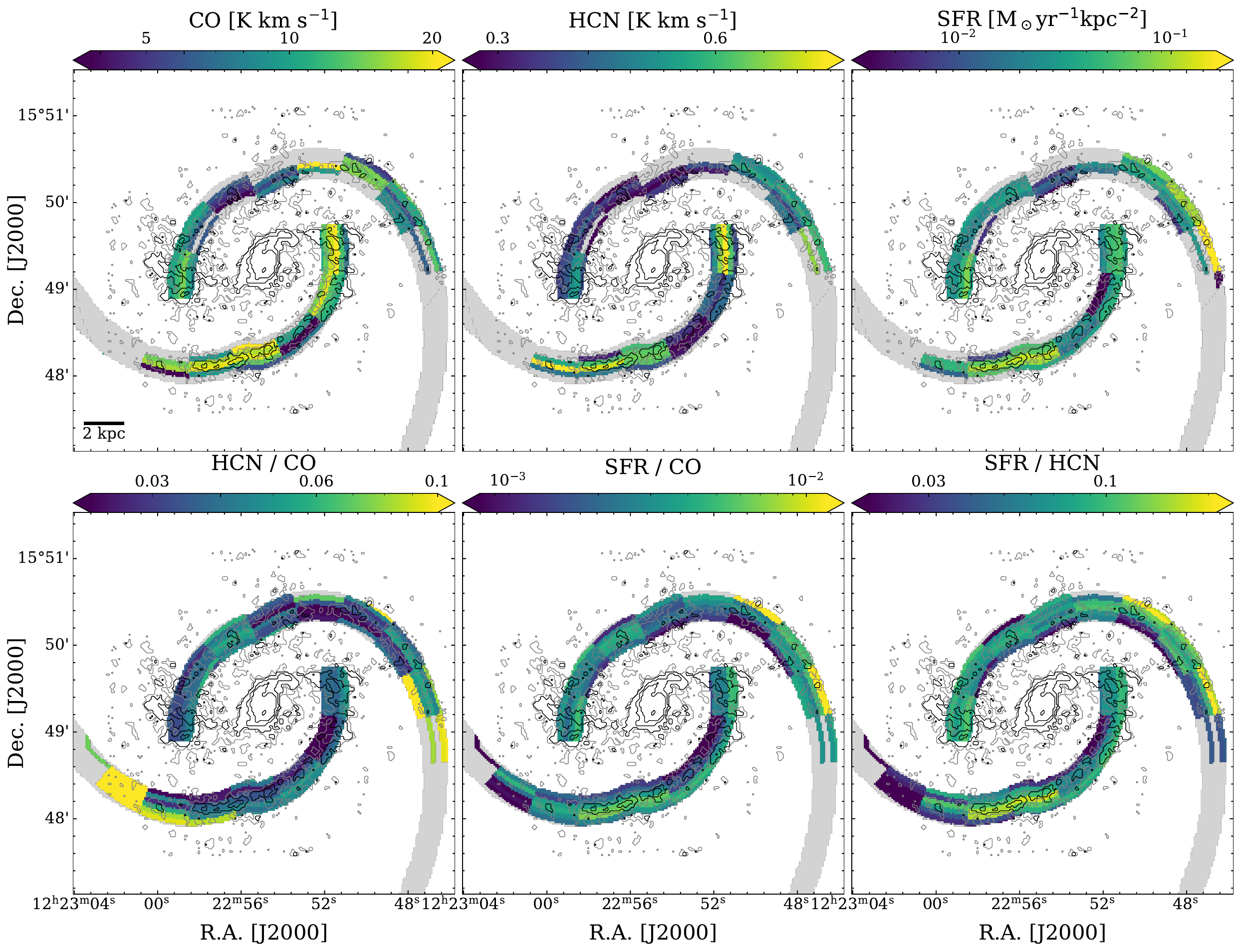}
    \caption{Rigatoni plot NGC\,4321: Maps of NGC\,4321 displaying the medians binned over $\Delta d$-intervals of $\SI{270}{\parsec}$ in accordance with the HCN beam size computed over $\sim \SI{2.5}{\kilo\parsec}$ wide phase bins. Top row: Distributions of the median integrated intensities of CO, HCN, and SFR (from left to right), masked with a S/N = 3 cut based on HCN. Bottom row: Ratios HCN/CO, SFR/CO, and SFR/HCN, based on a S/N = 1 masking based on HCN. The black contours show an HCN S/N of at least 3, 10, and 30, the grey contours additionally of at least 1.}
    \label{fig:Rigatoni_NGC4321}
\end{figure*}

\begin{figure*}
    \centering
    \includegraphics[width=0.7\linewidth]{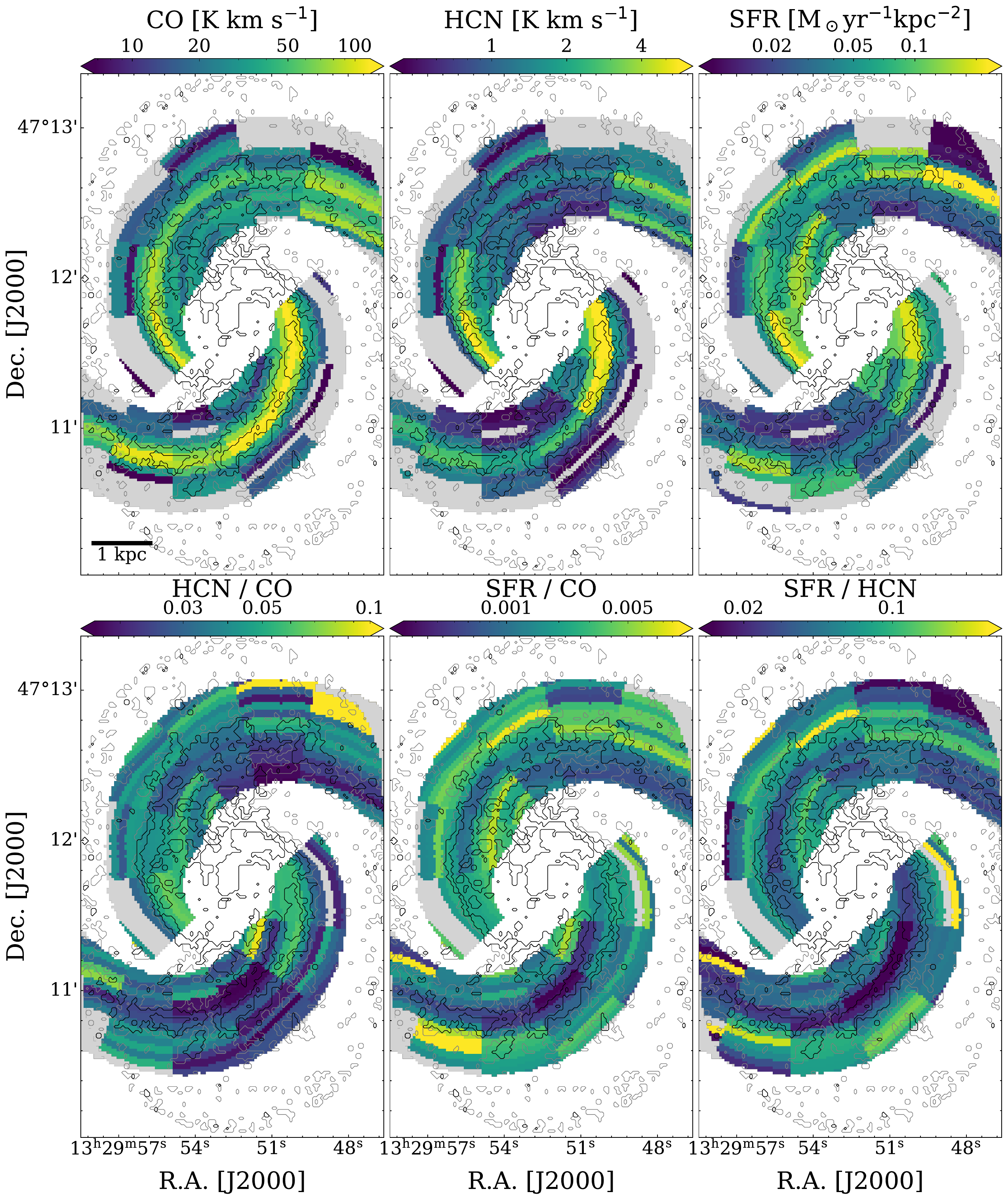}
    \caption{Rigatoni plot M51: Maps of M51 displaying the medians binned over $\Delta d$-intervals of $\SI{125}{\parsec}$ in accordance with the HCN beam size computed over $\sim \SI{1}{\kilo\parsec}$ wide phase bins. Top row: Distributions of CO, HCN, and SFR, masked with a S/N = 3 cut based on HCN. Bottom row: Ratios HCN/CO, SFR/CO, and SFR/HCN, based on a S/N = 1 masking based on HCN. The black contours indicate an HCN S/N of at least 3, 10 and 30, while the grey contours additionally show regions with S/N of at least 1.}
    \label{fig:Rigatoni_M51}
\end{figure*}

Similar to the Linguine plots in Figs.~\ref{fig:Linguine_NGC4321} and~\ref{fig:Linguine_M51}, Figs.~\ref{fig:Rigatoni_NGC4321} and~\ref{fig:Rigatoni_M51} display maps of the median integrated intensities of CO, HCN, and SFR and their ratios computed over the phase bins in NGC\,4321 and M51, respectively. The median trends are displayed as a function of the distance to the spiral arm spine ($\Delta d$) in Figs.~\ref{fig:NGC4321_ratio_delta_d_phase} and~\ref{fig:M51_ratio_delta_d_phase}.

\end{appendix}

\end{document}